\documentclass[iop]{emulateapj}
\usepackage{amsmath}
\usepackage{amssymb}

\newcommand{\boo}{Bo\"otes~I}
\begin{document}
\shorttitle{\sc Accurate kinematics in the Bootes-I dSph}
\shortauthors{Koposov, Gilmore, Walker et al.}

\title{Accurate Stellar Kinematics at Faint Magnitudes: application to
  the Bo\"otes~I dwarf spheroidal galaxy\footnote{Based on observations collected at the European
    Organisation for Astronomical Research in the Southern Hemisphere,
    Chile (proposal P82.182.B-0372, PI: G Gilmore)}}

\author{Sergey E. Koposov\altaffilmark{1,2},
  G. Gilmore\altaffilmark{1}, M. G. Walker\altaffilmark{1,3},
  V. Belokurov\altaffilmark{1}, N.Wyn Evans\altaffilmark{1},
  M. Fellhauer\altaffilmark{4}, W. Gieren\altaffilmark{4},
  D. Geisler\altaffilmark{4}, L. Monaco\altaffilmark{4.5},
  J.E. Norris\altaffilmark{6}, S. Okamoto\altaffilmark{1},
  J. Pe\~narrubia\altaffilmark{1}, M. Wilkinson\altaffilmark{7}, 
R.F.G. Wyse\altaffilmark{8},  D.B. Zucker\altaffilmark{9,10}} 
\altaffiltext{1}{Institute of
  Astronomy, Madingley Road, Cambridge CB3 0HA, UK}
\altaffiltext{2}{Sternberg Astronomical Institute, Universitetskiy
  pr. 13, 119992 Moscow, Russia} 
\altaffiltext{3}{Harvard-Smithsonian
  Center for Astrophysics, Cambridge, MA 02138, USA}
\altaffiltext{4}{Departamento de Astronomia, Universidad de Concepcion, Chile} 
\altaffiltext{5}{ESO, Chile} 
\altaffiltext{6}{Research
  School of Astronomy \& Astrophysics, The Australian National
  University, Mount Stromlo Observatory, Cotter Road, Weston, ACT
  2611, Australia}
 \altaffiltext{7}{Dept of Physics and Astronomy, University of
 Leicester, Leics, UK}
\altaffiltext{8}{The Johns Hopkins University,
  Department of Physics \& Astronomy, 3900 N.~Charles Street,
  Baltimore, MD 21218, USA} 
\altaffiltext{9}{Department of Physics and Astronomy,  Macquarie University, 
North Ryde, NSW 2109, Australia}
\altaffiltext{10}{Australian Astronomical Observatory, PO Box 296, Epping, NSW
  1710, Australia}
\begin{abstract}
  We develop, implement and characterise an enhanced data reduction
  approach which delivers precise, accurate, radial velocities from
  moderate resolution spectroscopy with the fibre-fed
  VLT/FLAMES+GIRAFFE facility. This facility, with appropriate care,
  delivers radial velocities adequate to resolve the intrinsic
  velocity dispersions of the very faint dSph dwarf galaxies. Importantly,
  repeated measurements let us reliably calibrate our individual
  velocity errors ($0.2 \leq \delta_V\leq 5$ km s$^{-1}$) and directly
  detect stars with variable radial velocities. 
  We show, by application to the \boo \, dwarf spheroidal, that the intrinsic
  velocity dispersion of this system is significantly below 6.5\,km/s reported
by  previous studies. Our data favor a two-population model of \boo, consisting
  of a majority `cold'
  stellar component, with velocity dispersion $2.4^{+0.9}_{-0.5}$\,km/s,
  and a minority `hot' stellar component, with velocity dispersion
  $\sim 9$\,km/s, although we can not completely rule out a single component
  distribution with velocity dispersion $4.6^{0.8}_{-0.6}$\,km/s. We speculate
  this complex velocity distribution
  actually reflects the distribution of velocity anisotropy in \boo,
  which is a measure of its formation processes.
\end{abstract}


\section{Introduction}

There is continuing interest in analysis of the number, nature,
masses, and evolutionary histories of the dwarf spheroidal (dSph)
satellite galaxies, found in moderate numbers around both the Milky
Way Galaxy and M31 in the Local Group.  They have typical half-light
radii greater than 100pc, low surface brightnesses ($\sim$ 25-30 mag/sq.
arcsec), and central velocity dispersions of several km/s, implying that the
luminous component is embedded in a dominant extended dark matter halo
($M/L_V\sim 10-100 M/L_{V,\odot}$; \citep{mateo98,gilmore07}).  Their
chemical abundances are low in the mean, correlate with dSph system
luminosity, show real intrinsic dispersion, and have chemical element
ratios systematically different from those of Galactic halo field
stars over the metallicity range covering most Galactic halo stars
\citep{kirby09,norris10,tolstoy09}. Their stellar populations differ
systematically from those in the Galactic field. All the astrophysical
evidence shows they are the oldest surviving bound systems, probably
forming very early from purely primordial gas, e.g., \citet{norris10}. How
do they relate to the very large numbers of surviving dark-matter
halos predicted by standard structure formation models?
 
The most luminous dSphs around the Milky Way, the `classical' dSphs,
were discovered photometrically through the second half of the twentieth
century, apart from the nearest and largest, the Sgr dSph, which was
discovered in position-velocity-photometry phase space
\citep{IGI94,IGI95}.  More recently, following availability of the
large-area photometric data from the Sloan Digital Sky Survey
\citep[SDSS;][]{york00}, a three-times larger sample of dSph has been
discovered, primarily of dSphs with lower intrinsic luminosities,
extending to the `ultra-faint' dSphs
\citep{willman05,belokurov06,zucker06,zucker06b,belokurov07,walsh07,belokurov08,
belokurov09}. These objects have extremely low surface brightnesses
(down to 30 mag/arcsec$^2$) and low luminosities (down to $L_V\sim
10^3 L_{V,\odot}$), such that the presence of a single giant star can
substantially affect the luminosity of the entire
galaxy \citep{martin08}.

Knowledge of the numbers and masses of dwarf satellites remains
crucial for understanding local galaxy formation, since
high-resolution cosmological N-body simulations of galaxy formation
generally predict a number of dark matter `sub-haloes' that is still
an order of magnitude larger than the number counted in and
extrapolated from observations (the ``Missing Satellites
Problem''\citep{klypin99,moore99}).  Considerable efforts have been
expended in simulating dwarf galaxy formation
\citep{barkana09,somerville02,benson03,ricotti10}, attempting to
lessen the tension between predictions from simulations and the
observations \citep{koposov09,maccio10,cooper10}.  Modelling the
census of $\sim 25$ known Milky Way dSphs for survey incompleteness
\citep{koposov08,walsh09}, one can estimate that the number of such
galaxies within the halo of the Milky Way could be several hundred,
although with a significant error-bar
\citep{tollerud08,koposov09}. Nearly 50 dSph are now known around M31,
where the whole area has been studied, but not to such low
luminosities. Empirical constraints on the dark halo masses  of these objects
are required in order to say whether these discoveries represent a step toward
consistency with standard cosmological models.

The structure of the paper is as follows: we first outline our
specific motivation, deriving reliable velocity dispersions in faint
dSph galaxies. We then introduce the \boo \, galaxy and our
observational approach, which is designed to test the reliability with
which we can derive accurate radial velocities of faint stars. In
section~4 we describe our data reduction procedures, which are
developed to ensure use of the full information content in the raw
data. In Section~\ref{sec:spec_fitting} we describe our enhanced
procedure for fitting the stellar spectra and deriving precise,
accurate radial velocities, as well as our approach to understanding
the errors. In section~6 we quantify how we detect radial velocity
variations. Section~7 compares our results with available literature
studies. In Section~\ref{sec:vel_disp} we describe our statistical
methodology to determine velocity dispersions from kinematic data, and
apply this to the \boo \, dwarf galaxy. In
Section~\ref{sec:conclusion} we conclude our study with some
discussion of the astrophysical implications.

\section{Measuring velocity dispersions in dSph galaxies}

Several factors complicate determinations of dSph galaxy masses.  The most
basic stems from the fact that we can measure only line-of-sight
velocities of dSph stars.  As a result, any estimation of a dSph's
total mass must contend with the fundamental degeneracy between mass and
anisotropy in the velocity distribution \citep[e.g.,][]{wilkinson2002,kleyna02}.
The second basic constraint, as more generally,
is that we very rarely see an outer declining dispersion profile (or
rotation curve), allowing determination of a ``total'' mass.  A third
key limit comes from having to suppose the dSph can be considered in
dynamic equilibrium - an assumption whose validity is certainly not
obvious for those extreme systems currently within a few tens of kpc
of the Galactic centre.  Accepting the equilibrium assumption, and
using just radial velocities, several recent studies
showed 
that the most robust mass it is
possible to estimate for faint dwarfs is the mass enclosed within the
half light radius \citep{pen2008,walker09,wolf10}.
However, estimates of even this quantity meet with
observational hurdles: one is that the ultra-faint dwarf galaxies have
very small number of stars ($\sim 10^{2-4}$), so for some objects the
number of targets available for spectroscopic observations with even
the largest telescopes is limited to a few tens \citep{simon07} of
predominantly faint stars, for which radial velocity errors may be of
the order of or larger than the velocity dispersion of the system.
Another difficulty of measuring the small velocity dispersions for the
faint dwarf galaxies using small number of stars is that it still
unclear how much they can be affected by the binary stars which must
be in the sample \citep{olszewski96,hargreaves96,minor10,mcc2010}. A perennial
limitation in all these studies is data
quality - are systematic and random uncertainties in the data
quantified as precisely as is claimed?

We have two ambitions in this study. The first is to improve our
knowledge of the \boo \, dSph galaxy. Its high scientific interest is
introduced in the first subsection below. Our second aim is to improve
our knowledge of how reliably and accurately we can measure radial
velocities of dSph member stars.
Our observations have
been designed to allow us to derive radial velocities with sufficient
precision, and with sufficiently well-known accuracy, to resolve the
intrinsic dispersion in a typical faint dSph galaxy, to properly
understand our measurement errors, to quantify the fraction of objects
with variable radial velocities, and hence to measure the intrinsic
velocity dispersion with reliable precision. This study provides a
template for how reliably, and how accurately, velocity dispersions
can really be determined in the faintest dSph galaxies. These
ultra-faint dSph have very few apparently bright member stars, so one
is forced to determine precision velocities for very faint stars from
moderate spectral resolution, moderate signal-noise ratio spectra,
with the relevant velocity accuracy being at most a few percent of one
pixel in the observed spectrum.  The stars additionally tend to be
very metal-poor, weakening absorption line strengths.

\begin{figure}
  \includegraphics[width=3.39in]{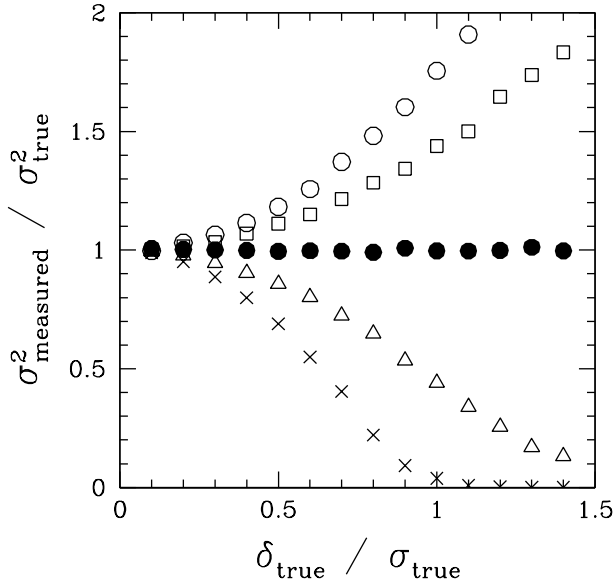}
  \caption{Accuracy of simulated velocity dispersion estimates as a
    function of the ratio of measurement errors to the true velocity
    dispersion.  Filled circles represent cases in which the
    measurement errors are known perfectly.  Open circles and open
    squares represent cases in which the errors used in the analysis
    are underestimated (optimistic) by a factor of 0.5 and 0.75,
    respectively.  Open triangles and crosses represent cases in which
    the adopted errors are overestimated (pessimistic) by a factor of
    1.25 and 1.5, respectively.}
  \label{fig:lowsigma}
\end{figure}

Kinematic studies of ultrafaint dSphs are especially challenging due to the
fact that the measured velocity dispersions the typical errors of individual
velocities, and the the expected contribution of binary stars are all similar,
at a few km/s. The previously studied SDSS dSphs have reported velocity
dispersions as small as $\sim 3$ km s$^{-1}$ \citep{simon07,martin07}.
\citet{mcc2010} demonstrate that for such cold dispersions, it is difficult to
disentangle contributions from random( i.e., reflective of the underlyin
gravitational potential ) and binary-orbital motions. It may even be that some
of the apparently small extreme ultra-faint  systems are not (currently) dark
matter dominated, but are dissolved star clusters, or tidal debris. In such
a case the velocity dispersion could be substantially lower than
$\sim 3$ km s$^{-1}$. Could we detect such an absence of dark matter?

In this situation it is critical to account accurately for velocity
errors---when propagated through calculations, optimistic/pessimistic
errors result in over-/under-estimates of the true velocity
dispersions.  That is, how important are the uncertainties on the
error bars? We explore the severity of this risk with some very
simple Monte Carlo simulations, in which we draw velocity samples of
100 stars (slightly larger than our present \boo \, sample), in order to
minimize sampling error and thereby to isolate bias due to inadequate
resolution) from a Gaussian distribution with true dispersion
$\sigma_{true}$, and then add to each velocity a ``true'' error drawn
from a second Gaussian with dispersion $\delta_{true}$.  In real
observations the true error is unknown; we therefore measure the
velocity dispersion of each artificial sample after adopting velocity
errors $k\delta_{true}$.  The factor $k$ is held constant for a given
sample and represents the accuracy of the adopted errors: $k=1$ if the
adopted errors are accurate; $k>1$ for unduly-pessimistic errors;
$k<1$ for over-optimistic errors.  We repeat the simulation for values
$k=0.5,0.75,1.0,1.25,1.5$, and
$\delta_{true}/\sigma_{true}=0.1,0.2,...,1.5$; the latter sequence
lets us examine accuracy as the true velocity dispersion dominates, or
is dominated by, the measurement errors.

Figure \ref{fig:lowsigma} displays the squared ratio of the
measured to the true velocity dispersion,
$\sigma_{measured}^2/\sigma_{true}^2$, as a function of
$\delta_{true}/\sigma_{true}$.  Plotted points represent average
values from $10^4$ trials at each accuracy level.  We recover the
required outcome, that if measurement errors are known perfectly,
intrinsic dispersions can be measured accurately, even if they are
dominated by measurement errors.  However, as the velocity error
becomes a significant fraction of the true dispersion, deductions
quickly become unreliable for even modestly misjudged errors.  There
is a particular danger of grossly underestimating velocity dispersions
with pessimistic errors: when $\delta_{true} \sim \sigma_{true}$, the
measured dispersion is only $65\%$ ($20\%$) of the true value if the
errors are overestimated by a factor of $k=1.25$($k=1.5$).

These results emphasize the caution necessary when measuring and
interpreting small velocity dispersions.  Two prerequisites for
obtaining reliable results are 1) sufficient resolution such that
typical velocity errors are smaller than the velocity dispersion,
and 2) accurate estimates of the velocity errors. We designed this study
with that lesson in mind.

\section{The \boo \, dwarf galaxy}

\boo \, was one of the first new dSphs discovered using the SDSS
photometric survey, by \citet{belokurov06}. \boo \, has a number of
interesting properties, but seems representative of the group of
newly-discovered intrinsically-faint dSph galaxies which are
(reasonably) far from the Galactic centre. Many recent studies are
available. Photometric and stellar population studies have been
completed by \citet{HWB08, dej08, SO10}; RRLyrae variability studies
by \citet{siegel06,dallora06}; Spectroscopic abundance studies by
\citet{munoz06, martin07, norris08, NYGW10, norris10}; and a HI 21cm
search by \citet{BF07}. Of specific relevance here, two
kinematic studies are available. \citet{munoz06} obtained
spectra of red giant branch (RGB) candidates over part of \boo,
selected from SDSS DR4. They used the WIYN telescope and the Hydra
multifiber spectrograph. Their data (for a 7 member-star sub-sample)
yielded a systemic velocity of $95.6\pm 3.4$\,km/s, and a central
velocity dispersion of $6.6\pm 2.3$\,km/s.  They derived $\sigma_0 = 9.0
\pm 2.2$\,km/s from a 12 member-star sample, a value which was adopted
in a later analysis by \citet{wolf10}. \citet{martin07}
observed candidate \boo \, red giants from SDSS (DR4) with
Keck/DEIMOS, finding a mean velocity of $99.9\pm 2.4$\,km/s, with
central velocity dispersion $\sigma = 6.5^{+2.1}_{-1.3}$\,km/s for their
final sample of 24 member stars with small ($\delta_v \lesssim10$\,km/s)
velocity uncertainties. From their kinematics, \citet{munoz06}
deduced a mass of $1.1^{+1.3}_{-0.5} \times 10^7 M_\odot$.  \citet{wolf10} adopt
the high velocity dispersion of $9.0 \pm 2.2$\,km/s, by
accepting the superset of the data of \citet{munoz06}. From
this they deduce a correspondingly higher mass.
 
In summary, \boo \, is some 65\,kpc ($m-M=19.07$) distant, has
absolute luminosity $M_{V,\odot} =-5.9$, is devoid of HI, has somewhat
elliptical (ellipticity =0.2) morphology, has a half-light radius of
240pc, has an apparently exclusively old metal-poor stellar
population, with a significant blue straggler sub-population, has a
mean [Fe/H] metallicity of -2.55, has an intrinsic abundance
dispersion with formal $\sigma = 0.45$, with a range in [Fe/H] of at
least 1.7dex, and has at least one member star which has
[Fe/H]$=-3.7$. 

\boo \, is ideal for a more detailed kinematic study. Observationally,
\boo \, has kinematic data available from two quite different
spectrographs (one fibre-coupled, one slit), so that both an internal
and an external test of the accuracy of our data is possible.  It is
at intermediate distance from the Galactic centre (70\,kpc
Galacto-centric), far enough that tidal effects need not be dominant, yet
close enough that observations far enough down the giant branch to
obtain useful statistical samples are feasible, in spite of its
(interestingly) low intrinsic luminosity. It has extremely low surface
density, so it is a clear test case for galaxy formation models, none of
which naturally creates very large, very low surface brightness,
galaxies. It has extremely low metallicity stars. The abundance
results suggest \boo \, is a survivor of a true primordial system,
quite likely forming prior to reionisation, enhancing its
interest. Its velocity dispersion is reported as being at least 6\,km/s,
a value which is feasible to measure to high accuracy. It seems to
have an extremely high apparent mass to light ratio, again enhancing
its intrinsic interest. It is visible from the VLT. The primary
challenge is that it is large on the sky.  Our present study,
involving a single FLAMES field, as with those in the literature,
covers only the central half-light radius of \boo.

\subsection{Target Selection}
\label{subsec:targets}

\begin{figure*}
\includegraphics{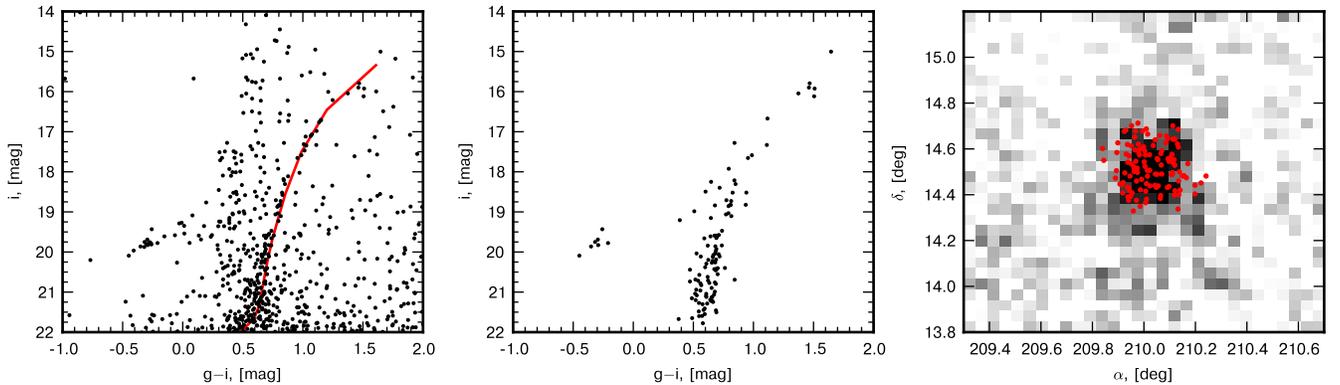}
\caption{{\it Left panel:} The color-magnitude diagram of all objects
classified as stars with good photometry in a 12armin square area
centred on \boo, from SDSS. {\bf The line shows the fiducial RGB sequence for 
the metal-poor globular cluster M92 \citep[from][]{clem08}, placed at the
distance of \boo\ dSph} {\it Middle panel:} The
color-magnitude diagram of candidate stars targeted for VLT
spectroscopy. {\it Right panel:} The density of
SDSS stars around the center of \boo \, is shown in grayscale, the
targeted stars are shown by red points. For comparison, the half-light
radius of \boo \, is 12.5arcmin.}
\label{fig:boo1_cmd}
\end{figure*}

For these observations, we selected target stars with a sufficiently
wide range of apparent magnitude that a single observation of the
brightest stars would provide similar signal-noise ratio to the final
signal-noise ratio of the fainter stars, integrated over 16 identical
single exposures. Thus, by comparing the actual repeatability of
velocities for the brighter stars, we could verify if the deduced
accuracy calculated for the faintest stars was reliable. Additionally,
by summing the brighter star exposures, we could readily test for any
systematic floor in delivered accuracy at some high signal-noise
value. We were fortunate to have access to the VLT FLAMES
spectrograph, since that, being fibre-fed, reduces substantially the
extra complexity of slit-centering errors, which unavoidably plagues
slit spectroscopy \citep{hargreaves94}.

We used the FLAMES spectrograph at the 8.2-m Kueyen (VLT/UT2)
telescope at Cerro Paranal, Chile, to acquire spectra of individual
stellar targets in \boo.  Observations took place in service mode
during 2009 February-March in fulfillment of ESO Programme 182.B-0372A
(PI: Gilmore).  For this programme we used FLAMES in UVES-Fibre
mode \citep{pasquini02}; that is, 130 fibres fed the medium-resolution
Giraffe spectrograph while eight additional fibres fed the
high-resolution UVES spectrograph.  The Giraffe spectra sample the
region $8180$\AA - $9375$\AA, in LR08 setup, including the prominent
CaII-triplet (CaT) absorption feature, with resolving power $R\sim
6500$.  The UVES spectra results will be described elsewehere (Monaco,
Gilmore et al, in prep).  Here we present results based primarily on
the Giraffe spectra.

In order to maximise the probability of observing \boo \, members, we
selected stellar targets based on colors and magnitudes of the stars. The
left-hand panel of Figure \ref{fig:boo1_cmd} displays the
colour-magnitude diagram (CMD) of all stars (including foreground)
within 12 arcmin of the center of \boo , from SDSS photometry and with
magnitudes corrected for extinction using the dust maps of
\citet{schlegel98}.  Recall that the half-light radius of \boo \, is
greater than 12arcmin - we are sampling only the very inner regions in
this (first) study.
 \boo's low luminosity presents a challenge to
spectroscopic studies, as the paucity of bright RGB candidates limits
sample size.
As the RGB of \boo\ together with a few blue
horizontal branch stars(BHB) are clearly visible on the left panel of Figure
\ref{fig:boo1_cmd}(see also Figure 1 from \cite{belokurov06}) the targets were
selected using a CMD mask covering the location of RGB and BHB stars.
The middle panel of Figure \ref{fig:boo1_cmd} plots the
CMD for our FLAMES targets, which include stars up to two magnitudes
fainter than the horizontal branch.  The UVES targets are all brighter
than $r\sim 18$, while the limiting magnitude of $r\sim 22$ for our
Giraffe targets represents a compromise between quantity and quality
of the spectra in our sample. As noted above, we retain several
(relatively) bright stars of low {\sl a priori} membership probability
as a key part of our internal quality checks. 

The second compromise involved the trade-off between target numbers
and adequate data to test our delivered velocity accuracy.
In order to obtain useful spectra for the faintest targets,
and to implement our test of achievable precision, we observe only a
single field with a single target configuration centered on \boo,
building signal-to-noise (S/N) ratios by repeated science
exposures. [The alternative would have been to reallocate those fibres
allocated to brighter target stars after one or a few integrations, to
increase observed numbers.] Our adopted strategy is well-suited for
service mode observations, given the excellent stability of the FLAMES
instrumentation.  Over the six weeks of our observing programme, we
obtained 16 individual science exposures, typically of 45 minutes
each.  The total exposure time on our field was 11.5hours.

\section{Data reduction}
\label{sec:data}

We first summarise the sequence of our data processing, before
providing a detailed description below:
\begin{itemize}
\item Default basic processing using the ESO pipeline (bias
  subtraction, flatfielding, wavelength calibration, extraction);
\item Wavelength recalibration of each extracted spectrum using sky emission lines;
\item Combining sky fibres for the determination of the master sky
  spectrum for each frame and subtracting from individual extracted
  objects;
\item Combining the individual frames into the co-added
  spectra. Co-added spectra are used then to reject cosmic rays from
  individual spectra;
\item Spectral fitting of the co-added spectra. We determine the
  best-fit template and approximate velocity;
\item Spectral fits of individual (not co-added) spectra using the
  best-fit template.

\end{itemize}

The initial data processing was done using the giraf-3.8.1 pipeline
provided by ESO with some modifications and bug fixes described
below. The important bug we discovered and fixed in the pipeline was
related to the computation of the variance spectra, which had been
incorrectly scaled by the pipeline. Another modification that we
applied to the pipeline was related to the extracted but not-rebinned
spectra. Our goal was to minimise the number of rebinning steps of the
data, so we needed to produce non-rebinned extracted spectra as well
as the corresponding wavelength solutions. As the original ESO
pipeline does not provide these, we modified our version of the
pipeline to output the necessary information, in un-rebinned pixel
space. All the further discussion of the data reduction will take into
account that each individual spectrum is in its native pixel space,
and so each spectrum is on a different wavelength grid. By contrast,
the common approach has all the data rebinned to a common wavelength
scale, and often a common dispersion, prior to all subsequent
anlayses.

\subsection{Wavelength calibration}\label{sec:wavecalib}

\begin{figure}
\includegraphics{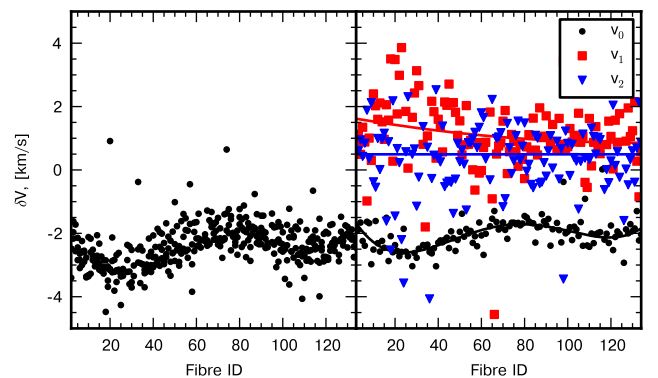}
\caption{{\it Left panel:} Velocity offsets from the ThAr arc
  wavelength fit, determined using the 3 sky lines 8310.7246\AA,
  8415.2422\AA, 8452.2656\AA, plotted for a single data frame, as a
  function of fibre ID number {\it Right panel:} The parameters of the
  polynomial wavelength corrections $v_0$,$v_1$,$v_2$ from
  Eq.~\ref{eq:wave_corr} as a function of Fibre ID. Different symbols
  show the measurements for individual fibres, while the lines show
  the polynomial fits to these measurements. Further explanantion is
  in the text.  }\label{fig:vel_offsets}
\end{figure}

The wavelength calibration was done using the standard Thorium-Argon
arc spectra, which were taken during the day, not in parallel with the
nighttime observations.  According to the GIRAFFE user manual this
calibration will deliver a precision limited only by spectral
resolution and signal-noise ratio, with a delivered instrumental floor
at very high spectral resolution of 30m/s.  After analysing the
extracted spectra, we noticed that the sky lines had systematic
velocity shifts of the order of 1$-$3\,km/s. The left panel of
Figure~\ref{fig:vel_offsets} shows these offsets measured using 3
isolated sky-lines (8310.7246\AA, 8415.2422\AA\ and 8452.2656\AA) as a
function of fibre ID for one of the GIRAFFE frames. A waving pattern
of velocity offsets versus the fibre IDs is clearly visible.  In order
to correct for these velocity offsets in each fibre we used all
(around 100) bright sky lines simultaneously. For each fibre we first
subtracted the continuum from the sky or from the stellar spectra,
then we fitted the residual spectrum in pixel space by a sum of n$\times$LSFs
(Line Spread Functions) at the locations given by the catalogue of sky lines
from \citet{uves03} after applying the polynomial correction to the
wavelengths.
\begin{equation}
\lambda=\tilde{\lambda}\,\left(1+\frac{1}{c}\,\left(v_0+v_1\,\frac{\tilde{
\lambda}-8400}{500} +
v_2\,\left(\frac{\tilde{\lambda}-8400}{500}\right)^2\right)\right)
\label{eq:wave_corr}
\end{equation}

Although the number of lines is high and the number of parameters is
high also, the fit may be done easily via sparse matrix operations. As
a result of this procedure we derive the best polynomial correction to
the spectra $v_0,v_1,v_2$. The results of determination of
$v_0,v_1,v_2$ for all the fibres on one particular GIRAFFE frame are
shown in the right panel of Figure~\ref{fig:vel_offsets}. Despite
considerable noise in the measured coefficients from fibre to fibre,
overall the coefficients vary smoothly with Fibre ID {\rm (Fibre ID is
enumerating fibers along the slit and on the CCD across the dispersion)}. In
order to
further reduce fibre-to fibre noise in the wavelength corrections we
fitted the $v_0,v_1,v_2$ coefficients by a set of low-order
polynomials (5th order for $v_0$, 2nd order for $v_1$ and 0th order
for $v_2$). The results of the fit are shown as coloured lines in the
right panel of Figure~\ref{fig:vel_offsets}). We use those curves to
determine the values of the parameters $v_0,v_1,v_2$ for each fibre,
and then substitute those values into Eq.~\ref{eq:wave_corr} in order
to determine the final wavelength solution for each fibre.

\subsection{Sky subtraction}

In each Giraffe frame we allocated 12 target fibres to sky, with sky
areas chosen to have no object detected by SDSS nearby. In order to
obtain an average sky spectrum for each frame, we first applied the
corrections from Sec.~\ref{sec:wavecalib}, then rebinned the sky spectra,
with twofold oversampling, from each individual fibre to the common
wavelength grid. We then multiplied each of the individual sky spectra
by a low-order(2) polynomial of wavelength to ensure that all rebinned
sky spectra had the same photon-count normalisation with wavelength,
and median-combined the rebinned sky spectra, creating an `average'
sky spectrum. For the error spectrum of the combined sky spectrum we
used the median absolute deviation (MAD), scaled by
1.4826$\sqrt{\frac{\pi}{2}}$/sqrt(number of sky spectra), following
\citet{mad93}.

The average sky was then subtracted from each individual fibre on the
frame. In order to perform this subtraction we had to rebin the
average sky spectrum back onto the original pixel wavelength grid of
each fibre. After subtracting the scaled sky spectrum from each fibre
we also added the variance of the mean sky spectrum to the variance
spectrum of each fibre to take into account the uncertainty of the sky
determination.

\subsection{Combining spectra}

Since each object in our sample was observed from 9 to 18 times, we
need to co-add the individual spectra. Although we did not use the
co-added spectra for the radial velocity determinations, we needed the
co-added spectra for certain specific tasks, for example, cosmic ray
rejection in individual exposures.

In order to combine repeated science spectra for a given star we first
made a zeropoint wavelength correction to correct for the
varying radial velocity component due to the Earth's motion, then
rebinned each individual sky-subtracted spectrum to a common wavelength
grid, as previously done for raw sky spectra. Then we followed a
procedure similar to that which we used in determination of the
average sky spectrum. Since our spectra were not flux corrected, before
combining the individual spectra we used 2nd order polynomials to
multiply all individual spectra, to ensure that each had the same
continuum level across the wavelength range. Then we median combined
these spectra. For variance spectra we used, as before, the median
absolute deviation(MAD), scaled by
1.4826$\sqrt{\frac{\pi}{2}}$/sqrt(number of observations).

Having the co-added spectra for each object allows us to identify bad
pixels/ cosmic rays/outliers in each individual (not co-added)
spectrum. In order to do that, we interpolated the co-added spectra to
the wavelength grid of each individual observation, and then we masked
out those pixels which were more than 4 sigma above or 6 sigma below
the median spectrum. Strictly speaking, that procedure may introduce
biases if the spectrum is highly variable from exposure to exposure --
in that cases the variable lines may be masked out, but we visually
checked all the spectra and did not see any inappropriately
masked spectral features.

\section{Velocity determination and spectral fitting}
\label{sec:spec_fitting}
The standard approach used for the measurement of stellar velocities
\citep{hargreaves94,koch07,simon07,walker07,baumgardt09,geha09,leaman09,koch09}
has been to cross-correlate against a template
spectrum \citep{simkin74,tonry79}. Although simple and computationally
fast, cross-correlation is known not to perform optimally
\citep{rix92,cappellari04}. In fact direct pixel-fitting methods have
been widely employed for more than a decade in spectroscopic studies
of unresolved stellar populations
\citep{rix92,cappellari04,chili07,koleva09}. Methods based on direct
pixel fitting provide more realistic error-bars and give a better way
to treat multiple templates and continuum levels.

In this section we briefly describe the pixel-fitting method we use,
which is similar to the ones described in
\citet{rix92,cappellari04,chili07,koleva09} but with a few
differences.

\subsection{Stellar library}
An important ingredient for direct pixel-fitting methods is the
library of template spectra. We decided to use the library of synthetic spectra
provided by
\citet{munari05}. The spectra in the library cover a large range of
stellar parameters $-2.5<$[Fe/H]$<0.5$, [$\alpha$/Fe] $=0$,0.4,
$3000<T_{\rm eff}/1K<80000$, $1.5<\log(g)<5$, $0<V_{\rm rot}/1km/s<150$. The
highest resolution spectra available from the Munari library are R $\sim$20000,
which is higher than the resolution of our data (R$\sim$6500), so the
templates can be easily downsampled to our resolution. While synthetic
spectral libraries have obvious limitations and we do not expect a
perfect match to the observed stars, the range of stellar parameters
covered in the library is large and is significantly better than what
we can achieve with stellar libraries of high resolution spectra for
real stars such as ELODIE \citep{prugniel07}; see also \citet{kirby08}. This is
particularly true given the extremely low
metallicities of the \boo\, member stars \citep{norris10}.

\subsection{Fitting synthetic templates to real spectra}
\begin{figure*}
  \includegraphics{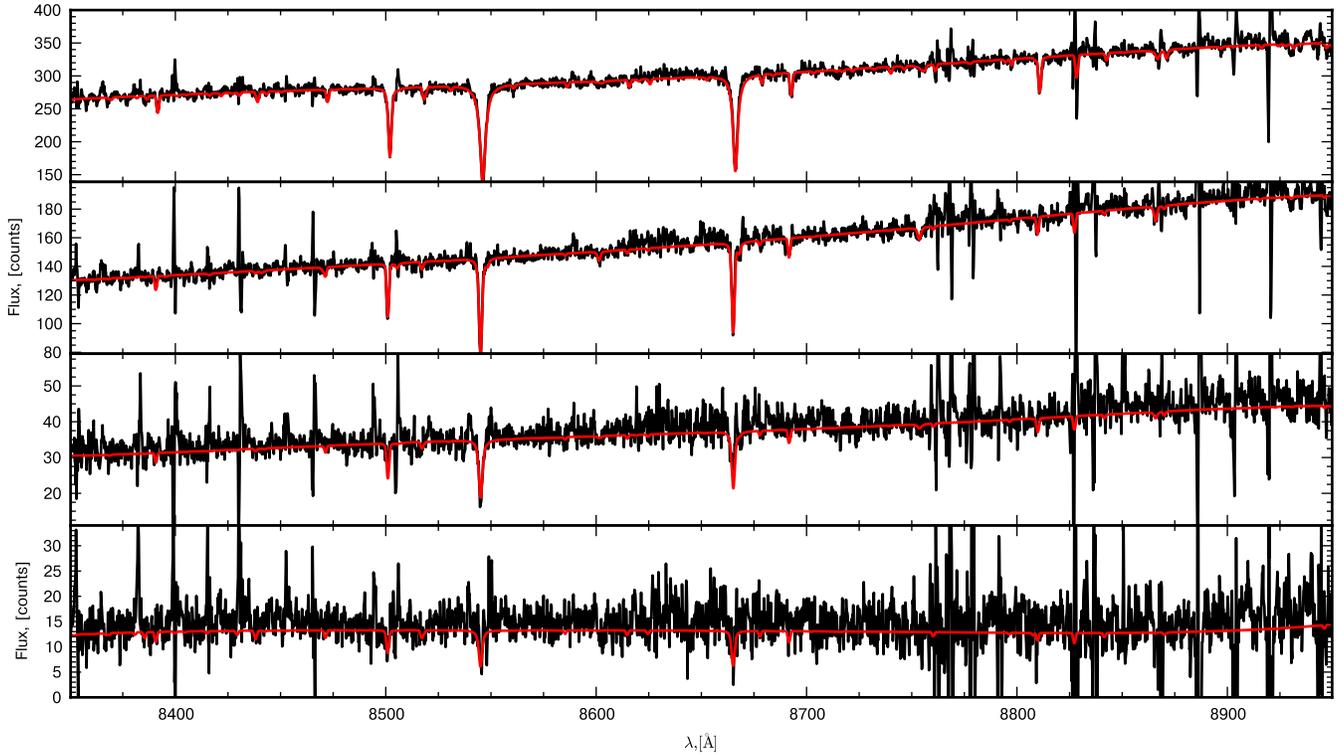}
  \caption{Examples of fitting synthetic model spectra to co-added
    spectra for several stars of different magnitudes (from top to
    bottom): SDSS J135922.59+143300.7 ($r=18.51$), SDSS
    J140001.53+142154.2 ($r=19.35$), SDSS J140002.44+142249.1
    ($r=20.62$), SDSS J135951.07+143049.8 ($r=21.52$). Black lines
    show the observed co-added spectra, while the red lines show the
    best fit model spectra. }
\label{fig:specfit}
\end{figure*}

Having the stellar template library we construct the model of each observed
spectrum as

\begin{equation}
Model(\lambda,i,v,\{p_j\})=P(\lambda)\cdot T_i\left(\lambda\,(1+\frac{v}{c})\right)
\end{equation}
where $T_i(\lambda)$ is the i-th template spectrum from the template
grid convolved with the appropriate Line Spread Function
$LSF(\lambda)$, $v$ is the radial velocity of the object and
$P(\lambda)=\sum\limits_{j=0}^{N-1} p_j\cdot\lambda^j$ is the
normalising polynomial of low degree, (N-1), which takes into account the
lack of flux calibration of our spectra, as well as any imprecision
in our location of the continuum of the template spectra.

Having the spectral model one can then compute the $\chi^2$ value by
summing scaled residuals over pixels.

\begin{equation} 
  \chi^2(i,v,\{p_j\})=\sum\limits_k\left(\frac{Sp_k-Model(\lambda_k , i , v, \{p_j\}) }{ESp_k}\right)^2
\label{eq:chisq_spec}
\end{equation}
where $Sp_k$ and $ESp_k$ are the observed spectra and variance spectra
respectively, and the $\lambda_k$ are the wavelengths of the pixels of
the extracted spectra. $Model(\lambda_k,...)$ is the evaluation of the
synthetic stellar model spectrum at the wavelengths of the individual
pixels. It is important that the original observed spectrum
is neither rebinned nor interpolated in any way, as this would lead to
correlated noise and information loss.

Equation~\ref{eq:chisq_spec} defines the $\chi^2$ or log-likelihood of
our model. In order to find the estimated radial velocity and the
best-matched template we need to minimise the $\chi^2$ with respect to
all relevant parameters. In fact we are not interested in values of
the coefficients of the normalizing polynomials $P(\lambda)$, so we
can marginalise over them and derive the joint probability
distribution of just the template identification $i$, and the radial
velocity, $v$. With a simple analytical computation we can derive that
probability distribution:

\begin{equation}
Prob(i,v)=(\det(M))^{-\frac{1}{2}} \exp\left(-\frac{1}{2}{\bf Y}^T\,M\,{\bf
Y}\right)
\label{eq:vel_prob}
\end{equation}
where ${\bf Y}$ is the vector having length N, such that :
\begin{equation}
{\bf Y}[j](i,v)=
	\sum\limits_k
	\frac{Sp_k\,T_i\left(\lambda_k(1+\frac{v}{c})\right)}
	{ESp_k}\,\lambda_k^j
\end{equation}
and M is a symmetric square NxN matrix, such that 
\begin{equation}
M[j_1,j_2](i,v)=\sum\limits_k
\frac{Sp_k^2\,T_i^2\left(\lambda_k(1+\frac{v}{c})\right)}{
ESp_k^2 } \,\lambda_k^{j_1+j_2}
\end{equation}

Equation~\ref{eq:vel_prob} describes the joint probability
distribution of an identified template and an associated target star
radial velocity $Prob(i,v)$. In practice we evaluate this probability
for a grid of plausible radial velocities $|v|<500$\,km/s (this is
effectively a uniform prior on radial velocity) and for  our grid
of templates (also assuming a uniform prior). This 2-dimensional
probability distribution can then be marginalised over template or
velocity. The marginalisation over template can be used to determine
the maximum a-posteriori (MAP) estimate of the velocity,
($\arg\max_v\, Prob(v)$), and of the velocity error.
The other marginalisation identifies
the best fitting template spectrum, and therefore an estimate of
stellar parameters.

There are a few important points to appreciate about the fitting
procedure used here. First, the data do not have to be rebinned to
either linear or logarithmic wavelength scales; therefore all the
information as well as the noise properties are preserved. Second, we
do not need to perform continuum subtraction from either the object or
the template, which is advantageous since that is always a poorly
defined task; instead we rely on the continuum shape from the
synthetic spectrum, with additional polynomial modifications
(cf \citet{koleva09} for a discussion of continuum fitting).

\begin{figure}
\includegraphics{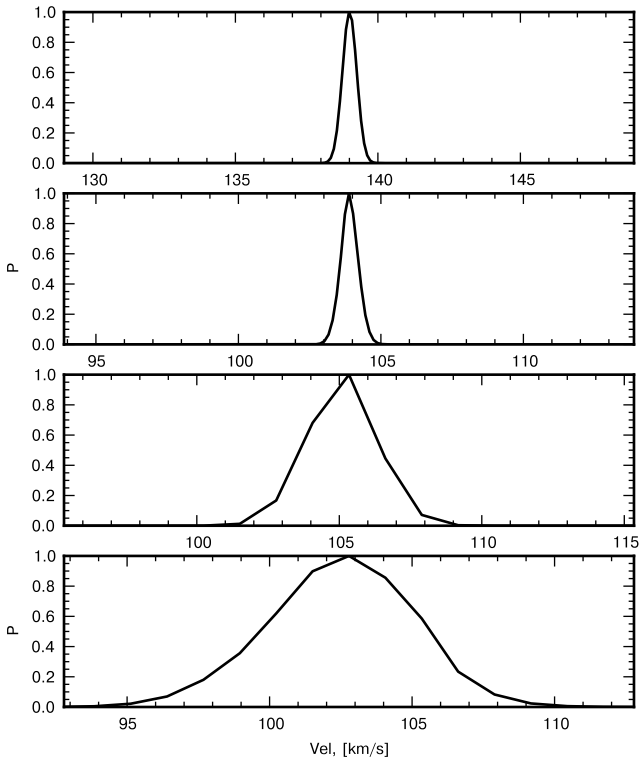}
\caption{Un-normalized velocity posterior probability distributions
  derived from the co-added spectra, for the same four stars as shown
  in Fig.~\ref{fig:specfit}.  It is clear that, especially at faint
  magnitudes/lower signal-noise, the velocity probability distribution
  is not Gaussian.}
\label{fig:vels_prob}
\end{figure}

Figure~\ref{fig:specfit} shows the result of the spectral fitting
procedure applied to the co-added spectra of four of the stars in the
\boo \, sample.  Figure~\ref{fig:vels_prob} shows
the resulting velocity probability distribution for the same four stars.
 The velocity precision ranges from
0.25\,km/s to 7\,km/s. For the brighter stars the probability
distribution of the velocity is approximately Gaussian, while for the
fainter star it is clearly asymmetric. For some of the faintest stars
in our sample the probability distribution is even multi-modal.

\subsection{Application to  \boo \, stellar spectra}

\begin{figure*}
 \includegraphics{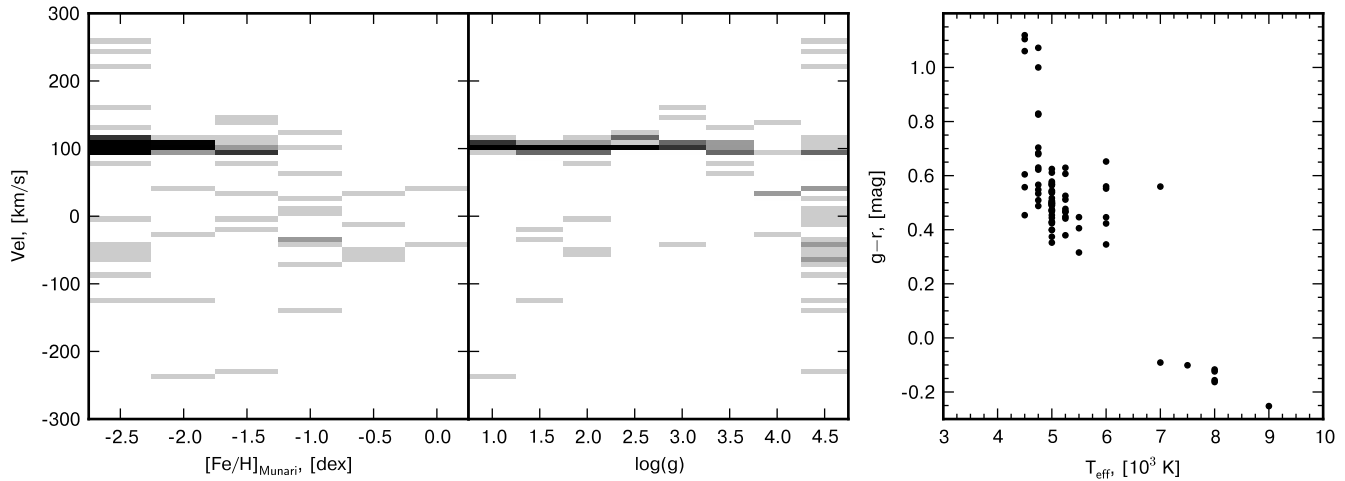}
 \caption{The stellar parameters determined from our synthetic spectrum
   template fit. The left panel shows the 2D histogram of radial
   velocities and metallicities of the best-fit templates for all
   observed stars. It is clear
   that the velocity peak at $\sim 100$\,km/s, the systemic velocity of \boo,
   occurs at $[Fe/H]\lesssim-2$dex, which is $\sim$ the metallicity of \boo,
   while the background (halo, thin and thick
   disk) has $[Fe/H]\gtrsim-1.5$. The middle panel shows that
most \boo\, stars have low
   surface gravities, as expected for giants, while the contaminating
   stars from the Milky Way foreground have almost exclusively
   $\log(g)\geq4.0$, as expected for dwarfs. The right panel shows the
   correlation of the g-r colour vs the effective temperature of the
   best-fit template. This correlation demonstrates that the
   effective temperatures are reasonably well determined. }
\label{fig:templparam}
\end{figure*}

For each star, as a first step we apply our fitting procedure to the
co-added spectra from each exposure. The purpose of using the co-added
spectra here is primarily to determine the best-fit template for each
object and the subsequent velocity estimate.  Although this fitting is
motivated by optimizing the radial velocity accuracy, and not
primarily to measure the stellar astrophysical parameters $T_{\rm eff}$,
$\log(g)$ and [Fe/H] from our template fit, the values of these
parameters determined from the best-fit templates are reasonable, as
illustrated by Figure~\ref{fig:templparam}.
 This Figure shows how the measured parameters
correlate with radial velocities and colors of the stars. 
The left panel of Figure~\ref{fig:templparam} shows the 2d distribution of 
radial velocities and metallicities of the stars. From the plot we see 
that the velocity peak at 100\,km/s
related to the \boo\ dwarf galaxy is produced by stars with
[Fe/H] $<-1.5$, which is what we expect from the metallicity of \boo \,
\citep{martin07,norris08}. In the middle panel of
Figure~\ref{fig:templparam} we show the 2D distribution of radial velocities
 and $\log(g)$. We see that the contamination (from
foreground Milky Way stars) typically has high surface gravity
$\log(g)\gtrsim4.0$ -- exactly what we expect from disk dwarfs. The
right panel of the figure shows the correlation between determined
effective temperature and (g$-$r) colour. Although of relatively low
significance a correlation still can clearly be seen. Overall we
conclude that the parameters from the template fit are reasonably well
determined using our fitting procedure and the adopted spectral
library.

As the next step we use the best fit templates as well as the initial
velocity estimates for the final fitting of each individual (i.e., not
co-added) observation of a given object in the same way as described
above. The velocity estimate from the fit to the co-added spectra,
plus or minus 50\,km/s, is used as a uniform prior on the radial
velocity. That is, we make an assumption that the radial velocity of a
given object in a single observation does not vary more than 50\,km/s
from the value measured from the co-added spectrum.  From this step we
obtain probability distributions of the velocities for each epoch
observation of each object. We can use these repeated velocity
measurements for several purposes: first in order to assess our
measurement errors, and second to check for possible
binarity/variability of the radial velocities (see
Section~\ref{sec:variab}). We end with a total sample of 112 stars for which we
have derived radial velocities.

\subsection{Checking the Derived Uncertainties}

\begin{figure}
  \includegraphics{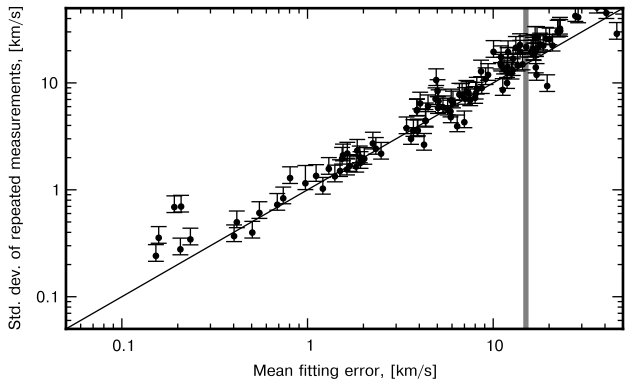}
  \caption{Comparison of the mean velocity error determined from the
    fitting procedure and the estimated standard deviation of the
    repeatedly measured velocities. Error bars show 68\% confidence
    limits for the estimated standard deviations. The solid black line
    shows the one-to-one relation. The grey line shows the approximate
    velocity where the one-to-one relation between the mean error and
    the estimated standard deviation is expected to fail due to
    non-Gaussianity of the probability distributions. 
}
\label{fig:errors_comparison}
\end{figure}

As described in the previous section, from our spectral fitting
procedure applied to either individual or co-added spectra we
determine the probability distribution of the radial velocity of a
given star $Prob(v)$, the MAP (maximum a-posteriori) estimate of the
velocity and the error of that velocity. One of the most important
checks on the validity of our results---as well as  a check on
the effectiveness of our reduction method---is the confirmation that the
velocity errors that we determine from individual observations of
specific (non-variable) stars are not systematically larger or smaller
than the scatter between individual repeated measurements. That is,
have we met our ambition of deriving correct and reliable uncertainties?

Figure~\ref{fig:errors_comparison} compares the standard deviation
determined from repeated measurements to the mean error determined by
our fitting procedure.
In this plot we expect to see a one-to-one correlation,
except for three cases: one, where there is intrinsic variability in
the radial velocity; two, where there is some remaining systematic
error which varies from observation to observation; or three, in the
case where measurement errors from individual exposures are
significantly larger than the one-pixel resolution of the spectrograph
($\sim$15\,km/s). In this latter case undersampling causes the
probability distribution of the velocities to become significantly
non-Gaussian, so the mean fitting error is an inadequate
representation of possible variation of the velocities.

With reassuringly few exceptions, the data points in
Figure~\ref{fig:errors_comparison} scatter about the one-to-one
line. There is a slight apparent systematic tendency for the data points to be
scattered more above the one-to-one line than below it, expected given
that the error-bars on the standard deviation are significantly
asymmetric.  The one-to-one relation extends in general down to a
precision of a few hundred m/s, confirming that our derived error-bars
are reliable down to 0.1$-$0.2\,km/s.  In the next section we show that
several points lie above the one-to-one line not because we have
underestimated our errors, but rather due to intrinsic variability of
their radial velocities.

Since we were somewhat, albeit pleasantly, surprised at obtaining
velocity precision substantially better than 1\,km/s, as confirmed by
repeated measurements, we checked whether this may be some artifact of
the algorithm due to the sky lines.  For example, if a systematically
poorly subtracted sky line affecting a specific wavelength caused
velocities to cluster near some fixed value, this situation could
mimic small variances in repeat observations. A piece of
circumstantial evidence against this hypothesis is that the nearly
one-to-one relation of the error-bars versus the standard deviation
from repeated measurements (Fig.~\ref{fig:errors_comparison}) contains
a set of stars with very different radial velocities, while it seems
unlikely that sky-line artifacts are present at all possible radial
velocities. The most compelling evidence against this hypothesis is
that, since we have radial velocity measurements spread over a month of
observations, the variation of barycentric corrections is roughly
3\,km/s, which is much larger than our claimed precision and
scatter between repeated measurements; this would not be possible if
our measurements were driven by some earth-velocity sky-features or
artifacts.

As a final cross-check of our radial velocity error-bars, we
compared them with the theoretical estimates of minimum possible
velocity uncertainty, based on Fisher-matrix like arguments
\citep{murphy08,griest10}. This comparison suggests that our
error-bars are not unrealistic, and are typically very close to that
minimum velocity uncertainty, being above it only for a few of the
faintest stars. This confirms that our velocity precision is not
unreasonable.

A last remark is that 
, while the random errors seem to be correctly determined for the brightest
stars with the highest precision($<$ 1\,km/s from individual exposures), there
is a possiblity that
the systematic errors due to template mismatch (e.g., stellar spectra looking
systematically different from spectra in our grid) are dominating the error
budget.

\section{Variability analysis}
\label{sec:variab}

The goal of this section is to use our velocities, and velocity
uncertainties, to estimate which of the observed stars
show significant intrinsic variability, so that we may remove those
stars from the analysis of the velocity distribution in \boo. We do
this by testing the plausibility of two hypotheses for each object,
the first hypothesis being that the object does not show significant
variation in radial velocities, and the second hypothesis being that the
object does show evidence for velocity variability with amplitude
greater than 1\,km/s.

For each star and each of N observations of each star our
spectroscopic fitting procedure provides us with the probability
distribution of the radial velocity $Prob_i(v)$ associated with that
observation. In general these probability distributions are not always
Gaussian, especially in the low signal to noise cases, so in the
following analysis we will avoid making any assumptions of
Gaussianity.

For each star we assume that there is a certain intrinsic distribution
of radial velocities $\Psi(v-v_{sys})$, where $v_{sys}$ is the
systemic velocity of the star. If the radial velocity of the source is
not varying then $\Psi(v-v_{sys})$ is a delta function. The likelihood
of the observations, assuming that they randomly sample $\Psi(v)$, can
then be written as
\begin{equation}
 L=\prod\limits_i^{N_{obs}} \int \Psi(v-v_{sys})\,Prob_i(v)\,dv.
\label{eq:likelihood_var}
\end{equation}

It is clear for example that if  all the individual $Prob_i(v)$ are
Gaussians, and if the object's velocity is assumed to be constant,
(ie, $\Psi(v-v_{sys})$
is a delta function), then the likelihood $L$ will be a Gaussian centered at the
weighted mean of the centers of the  Gaussians ($Prob_i(v)$), i.e. individual
velocity measurements.



However, we do not make these Gaussian assumptions, and can adopt a
more general approach.

In order to assess the velocity variability of a given object we
evaluate the likelihoods of our two hypotheses: H1, that the velocity
of the object does not change, that is
$\Psi(v-v_{sys})=\delta(v-v_{sys})$; and H2, that the velocity is
varying and the velocities are distributed uniformly between
$v_{sys}-s$ and $v_{sys}+s$. where $s$ is the measure of the scatter:
\begin{equation}
\Psi(v-v_{sys})=\begin{cases} \frac{1}{2\,s} & \mbox{if
  } |v-v_{sys}| < s\\ 0 & \mbox{if } |v-v_{sys}| \ge s \\ \end{cases}.
\end{equation}
For each hypothesis H1,H2 we can compute its likelihood using
Eq.~\ref{eq:likelihood_var}, and marginalise over the parameters
$v_{sys}$ and $s$, adopting uniform priors. The ratio of the
likelihoods of two hypotheses marginalised over the parameters becomes
a Bayes factor, B - which is a measure of the relative plausibility of
the two hypotheses \citep{jeffreys61,kass95,trotta07}. We calculate
this for all 112 stars in our final sample.

\begin{figure}
  \includegraphics{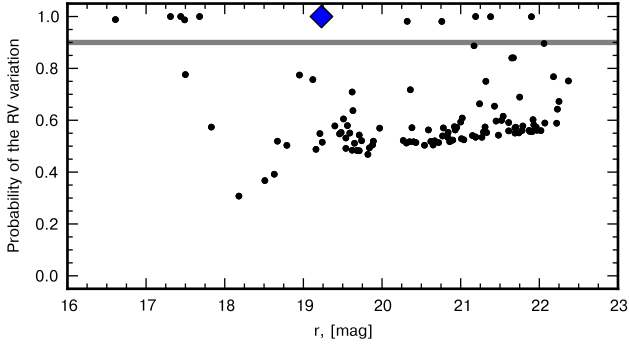}
  \caption{The probability of detected radial velocity variability for
    all 112 observed stars, computed using the Bayes factors for our
    two hypotheses: H1, there is no intrinsic variability in radial
    velocities; and H2, there is intrinsic variability with amplitude
    larger than 1\,km/s. The grey line shows the cut in Bayes factor,
    which is placed such that stars above the line favour the H2
    hypothesis, and are intrinsically variable. There are 11 such
    stars. The blue diamond indicates a known RR Lyrae star. }
\label{fig:bayes_factors}
\end{figure}

\begin{figure}
\includegraphics{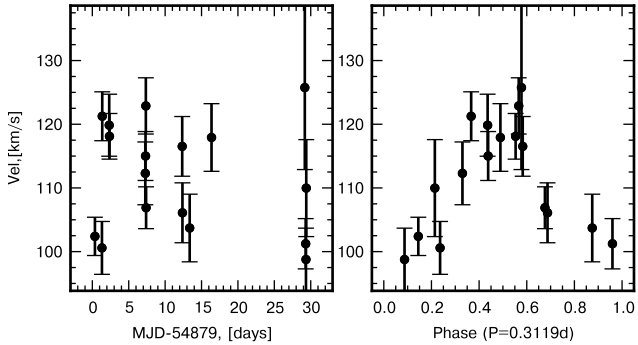} 
\caption{Radial velocity variation of the star SDSS
  J135951.33+143905.8, identified as an RR Lyrae variable by
  \citet{dallora06} and \citet{siegel06}. The left panel shows the
  observed radial velocities as a function of the Julian date, while
  the right panel shows the observations phased with the photometric period
  P=0.3119d from \citet{siegel06}.}
\label{fig:rrlyra_plot}
\end{figure}

\begin{figure}
\includegraphics{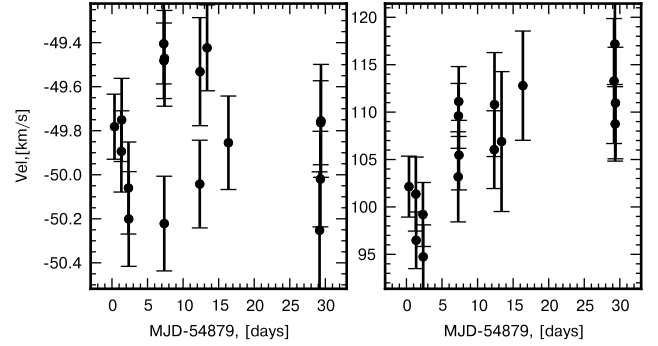}
\caption{Radial velocity variations for two stars in our
  sample, one indentifified as  variable, one not.  {\it 
    Left panel:} The radial velocity observations as a function of
  time for the fairly bright star in our sample, J135940.18+142428.0. Note
  the vertical scale. This is shown to illustrate the accuracy with
  which we are determining radial velocities. {\it right panel:}
  Radial velocity variation of star SDSS J140003.32+142851.4.}
\label{fig:variab_plot}
\end{figure}

Figure~\ref{fig:bayes_factors} shows the probabilities of variability
calculated for all 112 of our stars, derived from Bayes factors, as a
function of stellar apparent magnitude. The grey line shows the
location of the probability threshold \citep[see][]{kass95} which we
use to identify 11 stars with significant evidence for radial velocity
variation.  

The blue diamond point, located significantly above the grey line,
corresponds to one of our target stars. After (re-)discovering its
variability, we realised this star is a known RR Lyrae variable, from
published photometric studies \citep{dallora06,siegel06}. This
recovery does confirm that we can indeed correctly identify stars with
variable radial velocity.  Figure~\ref{fig:rrlyra_plot} shows the
observed (as a function of observation time), and phased using the
photometric period radial velocity variations for this RR Lyrae star.

 Figure~\ref{fig:variab_plot} illustrates the observed velocity data
 as a function of time for one bright, high signal-noise ratio star
 not detected to be velocity variable, and for another star
with variability Bayes factors above our variability detection threshold.

In short, the discussion above, and the evidence in the shown in these
figures, demonstrate that we are able to calculate a Bayes factor
probability that each observed star is consistent with being radial
velocity non-variable over the time in which we have observations.
While passing this test says little about much longer period velocity
variability, it does allow us reliably to identify stars for which we
have radial velocity data, but whose kinematics cannot be included in
a dynamical analysis at face value. Applying that criterion, we
restrict further analysis to 100 stars, with adequate velocity
measurements and error bars, and no evidence for velocity variability.

\section{Comparison with published kinematic studies of \boo}
\label{sec:comp}

Two observational kinematic studies of \boo \, \citep{munoz06,martin07}, and two
other analyses
\citep{wolf10,norris10,NYGW10}, have been published, providing
information which we may compare with our results.

\begin{figure*}
\includegraphics{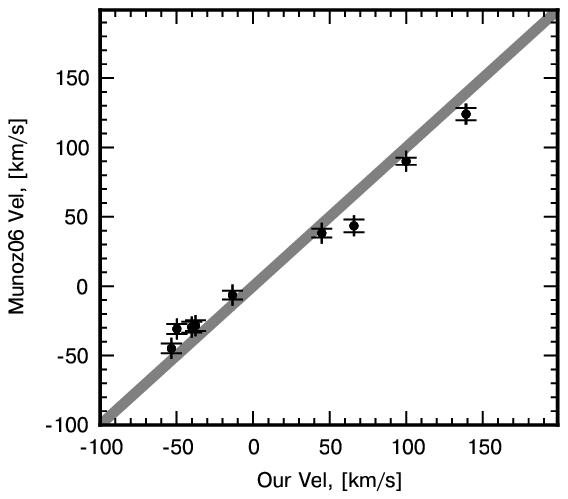}
\includegraphics{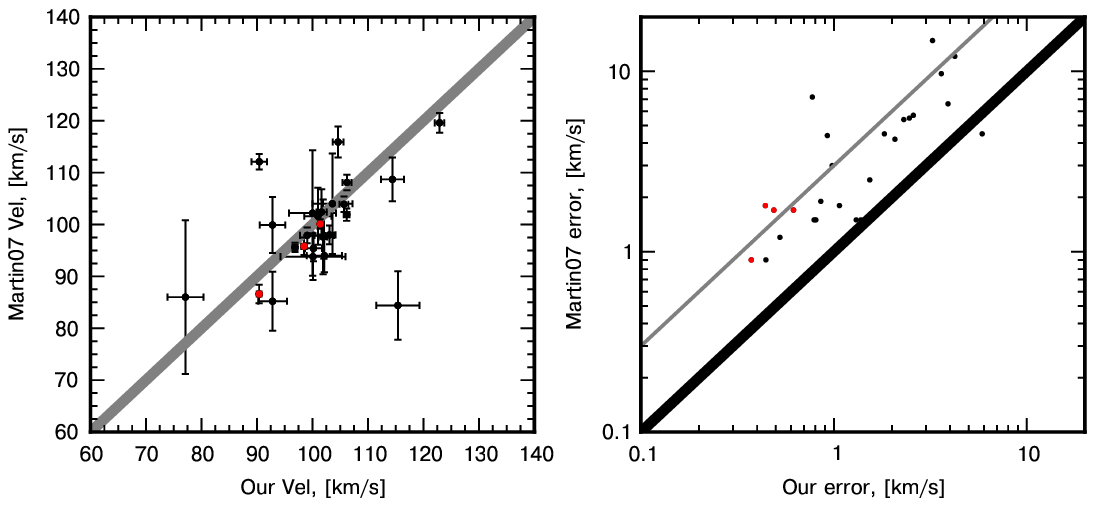}

\caption{{\it Left panel:} Comparison of our velocity measurements for
  9 stars in common with the velocity measurements of
  \citet{munoz06}. The line shows the expected one-to-one
  relation. The data sets are apparently not consistent withing the
  published errors, and suggest a velocity-dependent scale error.
{\it Middle panel:} Comparison of our velocity measurements for
  27 stars in common with the velocity measurements of
  \citet{martin07}. The line shows the expected one-to-one
  relation. The red points identify stars which may have variability
  in radial velocities (see Section~\ref{sec:variab}). {\it Right
    panel:} Comparison of our radial velocity errors with those of
  \cite{martin07}. The thick black line shows the one-to-one relation,
  while the grey line shows the $y=3\,x$ relation.
}
\label{fig:vel_compare}
\end{figure*}

\citet{munoz06} used the WIYN telescope and the Hydra
multifiber spectrograph to measure radial velocities for 58 candidate
member stars, all brighter than $g$=19. Considering the strength of
the Mgb absorption features, they classify fully 30 of these as
giants. Further considering position, velocity and line-strength, they
identify seven stars located within 10arcmin of the centre of \boo \,
to define the mean velocity and velocity dispersion of \boo, finding
a mean velocity $95.6\pm 3.4$\,km/s, and central velocity dispersion $6.6\pm
2.3$\,km/s.  With these values they further identify a total of 12 stars
as ``potential 3$\sigma$ members'', 11 being within a radius of
15arcmin of the centre of \boo, the 12th at 27arcmin distance (the
half-light radius is 12.5arcmin). These 12 candidate members provide a
higher mean velocity, $98.4\pm2.9$\,km/s, and a higher dispersion of
$\sigma_0 = 9.0 \pm 2.2$\,km/s; this higher dispersion, as the authors
note ``corroborating the apparent increase of the velocity dispersion
with radius''. The authors also note they have two further ``likely
high-velocity members''. The full set of 14 stars provides a velocity
dispersion of $\sigma_0 = 14.6 \pm 3.0$\,km/s, suggesting ``a possibly
dramatic increase of the Boo velocity dispersion with radius''. The
12-star sample, with its \boo \, velocity dispersion of $\sigma_0 =
9.0 \pm 2.2$\,km/s, is that adopted by \citet{wolf10} in their
dynamical analysis. We compare our results for 9 stars in common in the left
panel of Figure~\ref{fig:vel_compare}. Our results in this study are
inconsistent with the velocity dispersions reported by \citet{munoz06}, and we
rule out their suggested rapid radial increase in
velocity dispersion (see below).

\citet{martin07} observed 96 candidate \boo \, red giants from SDSS
(DR4) with Keck/DEIMOS. They identified a sample of 24 member stars,
each with a radial velocity determined with an accuracy smaller than
6\,km/s \citep[cf.][for details]{martin07}. From this sample they
find
a mean velocity of $99.9\pm 2.4$\,km/s, with central velocity dispersion
$\sigma = 6.5^{+2.1}_{-1.3}$\,km/s for their final sample of 24 stars
with small velocity uncertainties. We observed, by design, 27 stars in
common with the Martin et al. sample.

We compare our measurements for the 27 stars in common with the radial
velocity measurements by \citet{martin07} in the rightmost two panels of 
Figure~\ref{fig:vel_compare}.  The zero-points of our radial
velocity measurements are consistent. The mean velocities, for the
common 27-star member sample, are $\bar{V}_{keck}=99.6\pm1.7$\,km/s, and
$\bar{V}_{VLT,27}=101.2\pm2.0$\,km/s, in excellent agreement. The mean
velocity from our full sample is $\bar{V}_{VLT}=101.8\pm0.7$\,km/s.  The
right hand panel of Figure~\ref{fig:vel_compare} compares our
derived velocity errors with those reported by \citet{martin07}. It is
apparent our derived single-velocity errors are a factor of about
two smaller than those of \citet{martin07}, at high signal-noise ratios
in both studies.  We may also do a very crude check on the accuracy of
the quoted velocity uncertainties. Comparing stars in common with
\citet{martin07}, the velocity difference, in units normalised by the
quadrature sum of our quoted errors and those of \citet{martin07}, has
a sigma of 2.7. This is robust evidence the combined errors are
underestimated. The results of Figure~\ref{fig:errors_comparison}
suggest we have correctly calculated our uncertainties. In addition,
as we discuss in the section below, our derived velocity dispersion
for \boo \, is significantly below that published by other studies.  We
interpret this difference, in essence a difference between fibre-fed  and
slit spectrographs, as reflecting the inherent precision limits 
of velocity determination using multi-slit spectrographs.

For completeness, we note the studies of \boo \, by \citet{norris10} and
\citet{NYGW10}. \citet{norris10} obtained spectra of candidate RGB
members within a 1\,degree radius of \boo, using the AAT+AAOmega
facility. Their study was designed to identify candidate members of
\boo \, at large distances from the centre, for analysis of the chemical
abundance distribution, and to allow subsequent more detailed study for
kinematics, so they had, by design, relatively low velocity precision
(10\,km/s). Nonetheless, they identify candidate members out to 60\,arcmin
(5 half-light radii, cf their Fig~6) from the centre of
\boo. \citet{norris10} observed 5 stars in common with \citet{martin07}, and
report agreement in mean velocity within 3\,km/s and a
dispersion in velocity differences for those 5 stars of 2.3\,km/s (in
spite of their suggested 10\,km/s accuracy). They
further report work in preparation confirming one of their candidates
(Boo-980) to be an extremely metal-poor giant at 3.9\,half-light radii
from the centre of \boo, and with radial velocity $99.0\pm0.5$\,km/s.
\citet{NYGW10} report on a follow-up high dispersion UVES study of one
star identified by \citet{norris10}. This star, Boo-1137, is also extremely
metal-poor. Of relevance here, it lies
24arcmin (2 half-light radii) from the centre of \boo, and has radial
velocity $V_{uves}=99.1\pm0.5$\,km/s. Both these velocities are in
remarkable agreement with the mean velocity of the \boo \, system,
considering their large Bootes-centric distance, and hint at a low
velocity dispersion in the outer very metal-poor populations in \boo.

\section{The  velocity dispersion(s) in \boo}
\label{sec:vel_disp}
\begin{figure}
\includegraphics{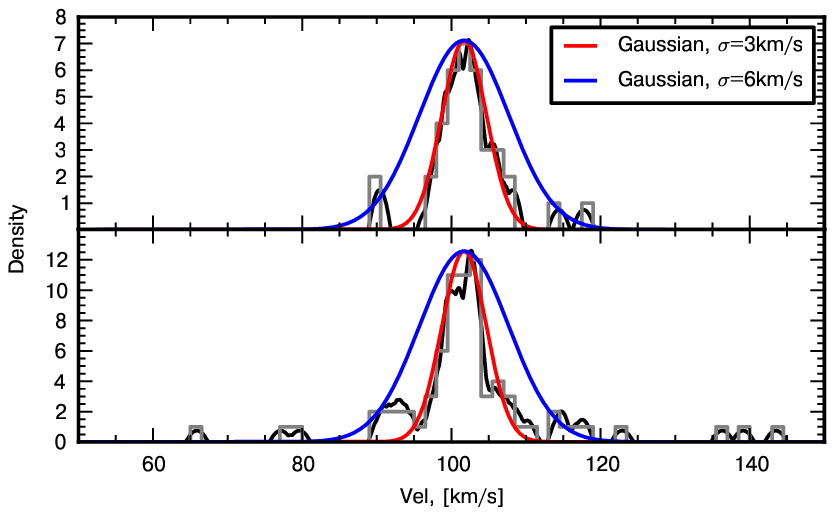}
\caption{The distribution of stellar velocities in \boo. The black
  line shows the distribution of velocities estimated using the
  Epanechnikov kernel\citep{epa69,wand95} with a bandwidth of 1.5\,km/s, the
grey
line shows a
  standard histogram with bin size of 1.5\,km/s. The red and blue
  lines are overplotted Gaussians with sigma of 3 and 6\,km/s;
  respectively; 6\,km/s is the smaller of the previously published
  determinations of the \boo \, internal velocity dispersion. The top
  panel shows the velocity distribution for our sample of 37 stars
  which are highly probable \boo \, members, i.e., those with [Fe/H]$<$-1.5,
  $\log(g)<$3.5, small velocity error $\sigma_v<$2.5\,km/s and no
  significant velocity variability. The bottom panel shows the
  velocity distribution for all 100 of our stars with non-varying
  radial velocity}
\label{fig:vel_hist}
\end{figure}
\begin{figure}
  \includegraphics{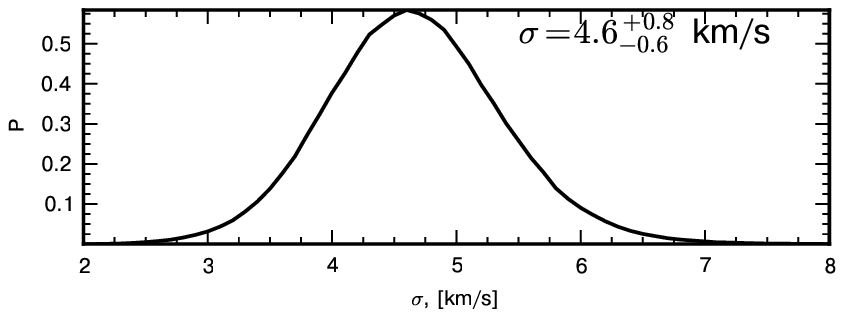}
  \includegraphics{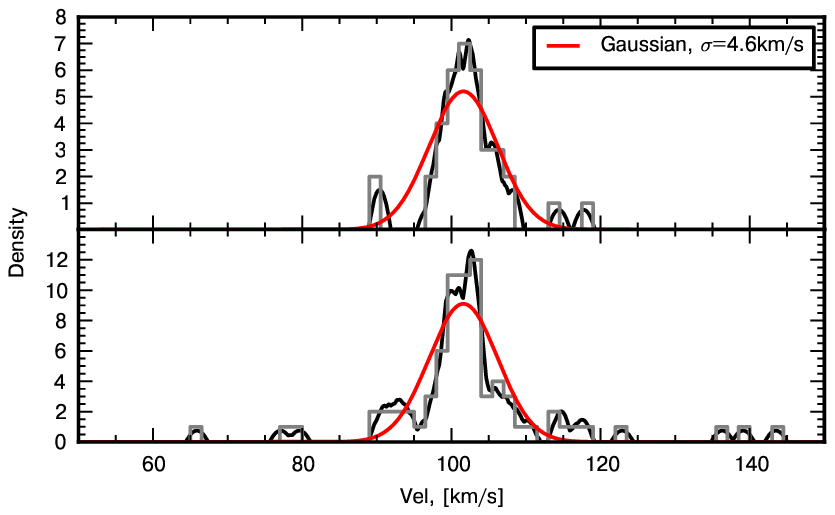}
  \caption{{\it Top panel}: The probability distribution of the internal \boo\,
    Gaussian velocity dispersion, determined from an MCMC fit to our
    velocity data for our full 100-star non-variable sample, when the
    velocity distribution is assumed to be consistent with a
    single Gaussian. {\it Middle panel}: the MCMC fit Gaussian, with
    dispersion 4.6\,km/s, overlaid on the kinematic data. {\it Lower panel}:
    to illustrate that the derived MCMC fit is robust to data
    selection, we show the derived Gaussian with dispersion 4.6\,km/s
    overlaid on the subset of 37 stars from figure~\ref{fig:vel_hist},
    those which are highly probable \boo \, members, ie those with
    [Fe/H]$<$-1.5, $\log(g)<$3.5, small velocity error
    $\sigma_v<$2.5\,km/s and no significant velocity variability}
\label{fig:vel_disp}
\end{figure}
\begin{figure}
\includegraphics{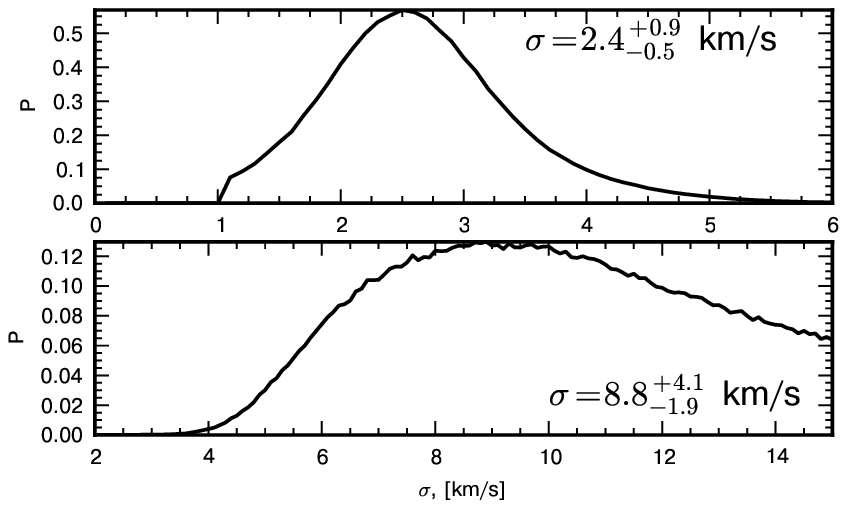}
\includegraphics{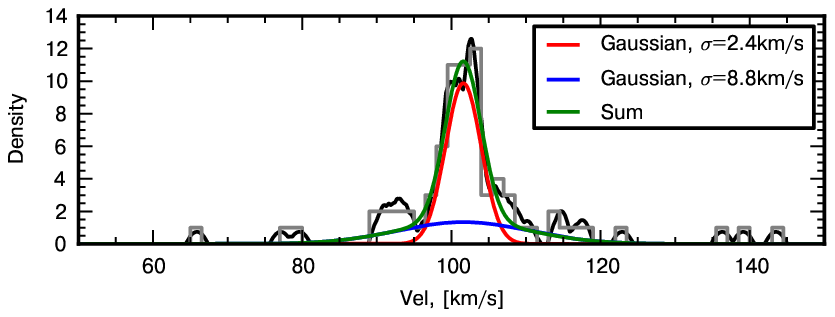}
\caption{{\it Top two panels}: The probability distributions of the internal
  \boo \, Gaussian velocity dispersion, determined from an MCMC fit to
  our 100-star non-variable velocity data, when the velocity
  distribution is assumed to be consistent with two Gaussians.  The
  top panel shows the probability distribution for the lower
  dispersion component, while the bottom panel shows the probability
  distribution for the higher dispersion component. The MCMC analysis
  allocated 70\% of the stars to the 2.4\,km/s dispersion component, and
  30\% of the stars to the 9\,km/s dispersion component. {\it Lower panel}:
  The corresponding two-Gaussian distribution overlaid on the
  kinematic data.}
\label{fig:vel_disp2}
\end{figure}

Equation~\ref{eq:likelihood_var} gives us a way to determine the best
velocity estimate from several velocity measurements. After removing
those stars which we suspect to be variable in radial velocity, we
model the remaining 100 star sample of velocities assuming that each
velocity is constant (i.e. $\Psi(v-v_{sys})$ is a delta-function). Thus we
derive the probability distribution of the systemic velocity for each star by
multiplying radial velocity probability distributions from separate
measurements $Prob(v)= \prod\limits_{observations} Prob_i(v)$. Those
probability distributions are in most cases very close to
Gaussians, so in the following analysis we assume that we have 
the velocity estimate and the Gaussian error-bar for each star. Table
\ref{tab:radvel} lists those velocities, and other information, for
each star.

Figure~\ref{fig:vel_hist} shows our resulting velocity distributions
for two different subsets of stars. The top panel shows the
distribution of velocities for the 37 stars which are likely members,
ie those with $\log(g)<3.5$, [Fe/H]$<-1.5$, not showing any variability
according to our Bayes factor criterion and with small errors of their
radial velocity measurement $\sigma_v<2.5$\,km/s. The bottom panel
shows the corresponding velocity distribution for all 100 non-velocity
variable stars in our sample. The velocity distributions illustrate 
two important points. First, the velocity peak due to \boo \, member
stars is quite evident. Second, it is clear that the velocity
dispersion of the stars in \boo \, is significantly smaller than 6\,km/s,
or 9\,km/s, the global dispersions measured previously
\citet{munoz06,martin07}. In the top panel especially, the
bulk of the distribution looks similar to a Gaussian with $\sim$ 3\,km/s
dispersion, while in the bottom panel the less restrictively selected sample
shows, in  addition to the low dispersion `core', rather pronounced
higher velocity tails. 

In order to assess the distribution of radial velocities we fit the
observed velocity distribution with two different models:
one where the velocity distribution in the galaxy is represented by a
single Gaussian, and the second with the velocity distribution being
the sum of two Gaussians with the same mean.  To perform these fits,
we follow the standard Bayesian approach.  We write down the
probability distribution as a function of the template metallicity,
template $\log(g)$ and radial velocity as the mixture of distributions for
the background (with fraction $f_{bg}$ and for \boo :
\begin{eqnarray}
P({\rm [\frac{Fe}{H}]}, \log(g), v) = f_{bg}\, P_{bg}{\rm
[\frac{Fe}{H}]}\, P_{bg}(\log(g))\,
 P_{bg}(v)+\nonumber\\
(1-f_{bg})\,P_{boo}{\rm [\frac{Fe}{H}]}\, P_{boo}(\log(g))\, P_{boo}(v)
\label{eq:prob_distr}
\end{eqnarray}
This technique minimizes the subjectivity involved in our selection,
for illustrative purposes, of the 37-star subsample noted above, where
we subjectively imposed an astrophysical prior, and which we restricted to the
small subset of highest precision data.

We assume that the probability distributions $P_{bg}{\rm
  [\frac{Fe}{H}]}$, $P_{bg}(\log(g))$, $P_{bg}(v)$, $P_{boo}{\rm
  [\frac{Fe}{H}]}$, $P_{bg}(\log(g))$ are Gaussians with different
means and standard deviations. For the radial velocity distribution of
stars in the dwarf $P_{boo}(v)$ we assume that it is either a single
Gaussian or the sum of two Gaussians with different dispersions but
the same mean. Having the probability distribution defined in
Eq.~\ref{eq:prob_distr} we perform a standard Markov Chain Monte Carlo
(MCMC) sampling of the available parameter space in order to determine
the posterior probability distribution for the parameters of
$P_{boo}(v)$. We use the standard Metropolis-Hastings algorithm (Metropolis
et al. 1953; Hastings 1970) implemented in the pymc package\citep{patil10}; 
see, e.g., \citet{neal93} for a review of the MCMC method.

The posterior probability distribution for the velocity dispersion of
\boo \, for the single Gaussian hypothesis is shown in
Figure~\ref{fig:vel_disp}. The velocity dispersion estimate is then
$4.6^{+0.8}_{-0.6}$\,km/s. This Gaussian dispersion is shown overlaid
on the kinematic data in the middle panel. To illustrate graphically
that our results are robust against sample selection, in the bottom
panel we show the same Gaussian distribution overlaid on the subset of
37 high-probability members with excellent data. The fit is
acceptable, but far from excellent.

If we make the assumption that the \boo \, velocity distribution
consists of two Gaussians [in addition to the fit to the background],
where one Gaussian has higher dispersion than the other, the posterior
probability distributions for the velocity dispersions of the two
components are those shown in Figure~\ref{fig:vel_disp2}. The velocity
dispersion of the lower dispersion component is then
$2.4^{+0.9}_{-0.5}$\,km/s, while the velocity dispersion of the other
component is not very well determined, but is around $9$\,km/s. The
fraction of stars belonging to the higher dispersion component
according to the MCMC fit is around 30\%. The corresponding Gaussian
distributions are overlaid on the kinematic data in the bottom
panel. The mean velocity from our full sample is
$\bar{V}_{VLT}=101.8\pm0.7$\,km/s.  In order to assess the probability
that the \boo \, velocity distribution is indeed described by two
Gaussians instead of one, we measure the likelihood ratio for these
two hypotheses, which gives $-2log(L_1/L_2)=8.06$.  This ratio
corresponds to a $\sim$98\% confidence of rejecting the single Gaussian
hypothesis.

Our data also provide direct limits on kinematic gradients in \boo. We
fit the MCMC modelling of the kinematics allowing for a linear
gradient in either RA (essentially the minor axis) or DEC (essentially
the major axis. Our formal limits on rotation are: minor axis,
$-4\pm9$\,km/s/deg; major axis $0\pm8$\,km/s/deg, recalling that our data
cover a radial range of 0.2deg.  We may also limit any radial change
in the dispersion. This is more difficult, since the minority
higher-dispersion component becomes dominated by sampling noise if the
sample is subdivided by radius. Hence we fit a single Gaussian
velocity dispersion model to the inner half (stars within 0.06deg of
the centre of \boo) and the outer half of the sample. The difference
in derived single-Gaussian dispersion is then (outer minus
inner)=$-2.1\pm1.3$\,km/s. This is not a significant result, but hints
that the dispersion may be {\it decreasing} with radius. That is, the
higher velocity dispersion component may (not significantly) be
somewhat more centrally concentrated than is the lower velocity
dispersion component. 

\section{Discussion and Conclusions}
\label{sec:conclusion}

In this paper we explain how we have developed and implemented a
thorough analysis of low-moderate spectral resolution ($R=6500$)
VLT/FLAMES stellar spectra taken in  the CaT wavelength region
near 860nm. We optimised a data reduction methodology which delivers
very accurate radial velocities, and very reliable uncertainties on
those radial velocities. We set up an optimised observational proof of
methodology, targeting faint candidate RGB stars in the very
metal-poor \boo \, dwarf spheroidal galaxy. By making 16 individual
repeat observations over six weeks, with a wide dynamic range
in each observational data set, we have reached several goals: most
importantly, we have been able to properly assess the errors of our
radial velocity measurements; our delivered radial velocity precision,
for faint extremely weak-lined stars, is better than 1\,km/s for the
combined exposures. Second, we have been able to identify and reject
stars that show significant radial velocity variability.

Comparing our derived velocity dispersion, and individual velocities,
with literature studies \citep{munoz06,martin07}, shows that previous
studies have substantially overestimated the velocity dispersion of
\boo. It is possible that earlier studies
underestimate their velocity errors, and hence overestimate the
velocity distribution which is deconvolved from those errors. Stable
fibre-fed spectrographs, including especially VLT+FLAMES, when
complemented with an appropriate observational strategy, and suitably
sophisticated data processing, are able to deliver precise, reliable
and accurate radial velocities with sufficient precision to resolve
the intrinsically very low velocity dispersions evident in the low
luminosity dSph satallite galaxies. By exploiting the spectrograph
stability to build integrations on times from days to years, we can
detect many radial velocity variables, whose unrecognised presence
would inflate erroneously a derived velocity dispersion.  Importantly,
from repeated observations, we can quantify reliably our velocity
accuracy. We are currently applying this observational technique to
several other dSph galaxies.

We have useful radial velocity measurements for 100 non-variable RGB
candidate stars, all within one half-light radius (12.5arcmin, 240pc)
of the centre of \boo. Approximately 60-70 stars are likely members
from our full sample.  Implementing a general MCMC analysis, which
includes separate \boo-member and background distribution functions of
our derived stellar parameters $\log(g)$, [Fe/H], and of the member
and background velocity distributions, we show that the distribution
function of stellar radial velocities in \boo \, which we measure can
be described in two ways. The less likely is that the distribution is
Gaussian, with a velocity dispersion of $4.6^{+0.8}_{-0.6}$\,km/s. The
more likely is that the distribution consists of two components: a
``colder'' component, containing 70\% of the member stars, which has a
projected radial velocity dispersion of $2.4^{+0.9}_{-0.5}$\,km/s, and
a ``hotter'' component, containing 30\% of the member stars, which has
a projected radial velocity dispersion of 9\,km/s. The data favor,
with 98\% confidence, the two component model.

Our data also provide direct limits on kinematic gradients in \boo. 
Our formal limits on rotation are: minor axis,
$-4\pm9$\,km/s/deg; major axis $0\pm8$\,km/s/deg. 
Similarly, we may limit any radial change in dispersion. This is more
complex, as fitting the full 2-component with a radial term becomes
sample-size limited. However, fitting a single Gaussian dispersion to
the inner half-radius and outer half-radius provides a formal,
statistically insignificant,  radial {\textit decrease} in dispersion of
$\-2.1\pm1.3$\,km/s. That is, there is a non-significant hint that the
apparently higher dispersion component is more centrally concentrated
than is the whole sample. 

We consider now the possibility that the higher velocity dispersion
component in \boo \, kinematics is an artifact of unresolved binaries
with variable radial velocities. This paper is based on the data taken
during a period of one month. While it definitely allows us to reject
stars with significantly variable radial velocities on hour to week
timescales, we do not of course identify all binaries. A hypothetical
substantial population of binaries with velocity amplitudes of order
10\,km/s, and periods much longer than a month, might produce spurious
wings in the radial velocity distribution which we would identify as a
hotter component.  Figure~\ref{fig:vel_compare} limits the
plausibility of this speculation: \citet{martin07} observed \boo \, in
2006, we observed \boo \, in 2009 - comparison of their velocity data
with ours for 27 stars in common identifies perhaps three stars with
velocity near that of \boo, and with sufficient velocity variability
to populate the distribution function wings. This is inconsistent with
the results of our MCMC analysis, namely that some 30\% of stars populate the
apparently higher velocity dispersion component. Thus taking the data of
\citet{martin07} at the face value, binary variability is unlikely to be the
cause of the higher-dispersion component we detect in our sample.

What can we deduce about the mass of \boo?  There has been recently a
flurry of studies on means to determine some useful mass parameter to
represent a dSph galaxy, where limited radial velocity data are
available, concentrated in the central regions. We noted above that
the most robust mass that can be estimated for faint dwarfs is
the mass enclosed within the half light radius \citep[for example, among
very many studies, ][]{walker09,wolf10}.
These several methods in
essence determine $M_{1/2} = \beta R_{1/2} \sigma_v^2$, where
$M_{1/2}$ is a characteristic mass inside the galaxy's half-light
radius $R_{1/2}$, $\sigma_v^2$ is the (Gaussian) dispersion of the
line of sight radial velocities, and $\beta$ is a factor of value
2.5/G to 3/G, with G the Newtonian constant. For a given object,
clearly this mass scales as the square of the velocity
dispersion. Equally clearly, it assumes there is a single well-defined
Gaussian dispersion.

For \boo \, we seem to have two dispersions, with no robust way
to associate a scale length with either. If we just assume that the
dominant low dispersion component is associated with the measured
half-light radius, then we can (very approximately) deduce an
associated mass within that radius of 240pc. It is perhaps more useful
to consider the range of determinations of the half-light mass
of \boo. \citet{wolf10} adopt a velocity dispersion for \boo \, of
9.0\,km/s, which they correspond to a mass  $M_{1/2} =
2.36.10^7M_\odot$, and $M/L=1700$. Fixing the geometric parameters,
but adopting the dispersion derived by \citet{martin07}, $\sigma_v
=6.5$\,km/s, corresponds to a mass lower by a factor of 1.9. Adopting our
low dispersion of  $\sigma_v = 2.4$\,km/s provides a mass lower than
that of \citet{wolf10} by a factor of 14, and a corresponding
$M/L=120$. The range of these numbers, depending entirely on the
quality of the kinematic data, clearly illustrates  both the need for
excellent quality data to allow useful study of the faint dSphs, and
the considerable caution which should be applied to extant analyses of
the masses of very low luminosity dSph galaxies.

\subsection{Two populations in \boo: some speculation}

Our data suggests that the \boo\ dwarf spheroidal has both cold and
hot stellar components. Although the data do not completely rule out a
single stellar component with velocity dispersion of 4.6\, km/s, that
scenario is not favored by our data. In order to understand the origin
of the two components we checked whether the components differ in
radial distribution or metallicity, but we were not able to find any
significant differences. This is not a strong constraint, given i)
that our data extend over only one half-light radius from the centre
of \boo, and ii) our metallicities are unable to resolve the extremely
low abundances of \boo \, members. \citet{norris10} show (their
fig~17) that the mean abundance of \boo \, is [Fe/H]$=-2.5$, with a tail
down below [Fe/H]$=-3.5$. These abundances are below the bottom range of
our template spectra, and beyond our ability to resolve in this study.
At face value two velocity components correspond to two scale
lengths. Might \boo \, have an extended higher dispersion component?
The available Subaru photometry (Okamoto, 2010) shows no indication of
an extended `envelope' structure, but is not a strong constraint
beyond 2 half-light radii, as the stellar surface density is extremely
low. We did note in Section~\ref{sec:comp} that radial velocities are
available for two extremely metal-poor members of \boo, at distances
of 2.0 and 3.9 half-light radii from the centre. Both velocities are
within 0.5\,km/s of the systemic velocity. While only two stars, they
are consistent with an extremely extended, very metal-poor, low
velocity dispersion component in \boo. These two stars were however
selected for detailed analysis because they were suspected of being
very metal-poor. One may not deduce that the more metal-poor stars in
\boo \, are the more extended. These two stars do however provide some
evidence against significant rotation or tidal warping of \boo, even
out to 1\,kpc. 

Perhaps more interesting, and consistent with a velocity
distribution function which is not a single component, is the radial
distribution of the 16 member stars with [Fe/H] abundances derived by,
and listed in Table 3 of, \citet{norris10}. If we divide the 16 stars
into equal inner and outer groups of eight stars based on radius from
the centre of \boo , the inner and outer groups have mean abundances
[Fe/H]=$-2.30 \pm 0.12$, and [Fe/H]=$-2.78 \pm 0.17$
respectively. These differ at the 2.4$\sigma$ level. 

An alternative interpretation is to question the validity of fitting
Gaussians to the radial velocity distribution function. In simple
  models \citep{gerhard93} strong deviations from a Gaussian
  distribution of radial velocities may arise from the velocity
  anisotropy of the stellar population. Given our lack of
understanding of how dSph galaxies form, there is no physical basis
for assuming a single isotropic velocity distribution. There are two
good reasons to doubt the single gaussian assumption for the very
faintest dSph. Firstly, just how does one populate an extremely
low-density distribution with scale length 250pc, out to apparently 4
scale lengths, 1000pc, with a velocity dispersion below 4\,km/s?
Dispersing a central star formation region in a very shallow (cored?)
dark matter potential may be feasible. Or it may not. Such a process
will inevitably generate very radially biased orbits. One alternative
speculation is to merge several star forming regions, none of which
need have been centrally located, during first formation of the dark
matter potential which we call \boo. This might well generate highly
tangentially biassed velocity distributions, where one might
anticipate the more tangentially-biased orbits to be more centrally
concentrated. Both processes might happen, so that the velocity
distribution may well be bi- or multi-modal in anisotropy, with
significant radial gradients in this anisotropy. The least likely
expectation is that the velocity distribution function is isotropic -
there is no obvious physical process which could generate such a
distribution function at such low densities and such large scales.

A plausibility argument of relevance to this speculation is that the
lowest luminosity dSphs are systematically less round than are more
luminous galaxies - although \boo \, itself has ellipticity 0.2, so is
relatively round. This flatness may be generated by an isotropic
kinematic distribution in a flattened potential, or equally by an
anisotropic kinematic distribution in a spherical potential. It is
often assumed that  dSph shapes correspond to the shapes of the
dominant dark matter distributions. Since one of the very few things
we suspect we know about the low luminosity dSph is that mass does not
follow light (more correctly, light does not trace mass, kinematics
do), the shape of the luminosity distribution may reflect kinematic
anistropy, not mass anisotropy. 
In that case radially anisotropic distributions
will appear in projected radial velocities as a distribution which is
more ``cuspy'' than is a Gaussian, while tangential anistropy will appear
more  extended than is a Gaussian \citep{gerhard93,bt08}. The
combination of these effects can look rather like our measured \boo \, velocity
distribution in Figure~\ref{fig:vel_disp2}. 

Thus, one may interpret our statistically preferred two-Gaussian
radial velocity distribution function in two ways. One option is that
there is a two-component structure in \boo, with future observations
at large Bootes-centric distances required to identify the
characteristics of the higher-velocity dispersion population, which is
a minority in the inner regions, and which must have a hugely extended
radial scale length. Alternatively, \boo \, has a velocity
distribution function which reflects its formation, and which is a
combination of a majority population with very radially-biased
orbits, and a minority population with very tangentially biased
orbits. Both are represented by a single, measured, radial scale
length. The more radially-biased population contains the most
metal-poor stars. 


If this speculation is valid, one anticipates that future precision
determination of the velocity distribution function in the most
flattened dSph galaxies (Hercules, $e=0.6$ and UMa~II $e=0.54$) will
find neither is consistent with being a single component Gaussian, and
will find a radially-variable shape for the distribution function of
radial velocities.

Most importantly, and our conclusion, it is evident that radial
velocity data of very high precision, and of extremely well-defined
accuracy, are necessary to make progress in defining the kinematics
and masses of the faintest dSph galaxies.

\acknowledgements{We thank Michael Siegel and Ricardo Mu{\~n}oz for providing us
with their
  photometric variability and kinematic data for \boo. Most of the data processing
  has been done using the python programming language and the following open
  source modules: numpy\footnote{\url{http://numpy.scipy.org}},
  scipy\footnote{\url{http://www.scipy.org}},
  matplotlib\footnote{\url{http://matplotlib.sf.net}},
  pypar\footnote{\url{http://code.google.com/p/pypar/}},
  ipython\footnote{\url{http://ipython.scipy.org}}\citep{ipython},
  astrolibpy\footnote{\url{http://code.google.com/p/astrolibpy/}} and
pyfits\footnote{\url{http://www.stsci.edu/resources/software\_hardware/pyfits}}
  (pyfits is a product of the Space Telescope Science Institute, which
  is operated by AURA for NASA).
  This research has also made use of NASA's Astrophysics Data System
  Bibliographic Services and TOPCAT
  software.\footnote{\url{http://www.starlink.ac.uk/topcat/}}\citep{taylor05}
  WG, DG and MF gratefully acknowledge financial
support for this work from the Chilean Center for Astrophysics FONDAP
15010003, and from the BASAL Center for Astrophysics and Associated 
Technologies CATA under grant PFB-06/2007. MGW is supported by NASA through
Hubble Fellowship grant HST-HF-51283, awarded by the Space Telescope Science
Institute, which is operated by the Association of Universities for Research in
Astronomy, Inc., for NASA, under contract NAS 5-26555.
}

\noindent{\it Facilities:} {ESO, VLT:UT2, VLT:FLAMES+GIRAFFE}

\clearpage
\begin{deluxetable}{ccccccccccccc}
\tabletypesize{\footnotesize}
\tablewidth{0pt}
\tablecaption{\label{tab:radvel}  }
\tablehead{ \colhead{ID}  &\colhead{Name}  &\colhead{$\alpha$}
  &\colhead{$\delta$}  &\colhead{$g$}  &\colhead{$r$}  &\colhead{RV}  
&\colhead{RV error}  &\colhead{P$_{var}$}  &\colhead{T$_{\rm eff}$}  
&\colhead{[Fe/H]}  &\colhead{$\log(g)$}  &\colhead{Bestflag}   \\ 
\\ \colhead{}  &\colhead{}  &\colhead{[deg]}  &\colhead{[deg]}
&\colhead{[mag]}  &\colhead{[mag]}  &\colhead{[$km/s$]}
&\colhead{[$km/s$]}  &\colhead{}  &\colhead{[K]}  &\colhead{}  &\colhead{}  &\colhead{}  }
\startdata
0&  J135921.36+143606.3&  209.8390&  14.6017&  17.8&  17.0&  -13.50&  0.10&  0.57&  4750&  -0.5&  4.5&   \\
1&  J135922.59+143300.7&  209.8442&  14.5502&  19.0&  18.5&  139.12&  0.35&  0.77&  5250&  -1.5&  4.0&   \\
2&  J135933.50+142821.6&  209.8896&  14.4727&  21.9&  21.5&  115.40&  3.91&  0.56&  5000&  -2.5&  2.5&   \\
3&  J135934.36+143017.1&  209.8932&  14.5048&  22.0&  21.5&  77.11&  3.25&  0.56&  4000&  -2.5&  3.5&   \\
4&  J135934.77+143503.4&  209.8949&  14.5843&  21.2&  20.7&  -228.29&  1.36&  0.54&  5000&  -1.5&  4.5&   \\
5&  J135935.66+143735.0&  209.8986&  14.6264&  20.0&  19.4&  -19.40&  0.44&  0.57&  4750&  -1.5&  1.5&   \\
6&  J135937.60+142647.6&  209.9067&  14.4466&  21.6&  21.1&  103.23&  2.54&  0.59&  5000&  -2.5&  1.0&   \\
7&  J135939.36+142638.4&  209.9140&  14.4440&  20.4&  19.9&  101.91&  0.77&  0.57&  5250&  -2.0&  2.5&  B\\
8&  J135940.18+142428.0&  209.9174&  14.4078&  17.5&  16.4&  -49.80&  0.05&  0.78&  4500&  -0.5&  4.5&   \\
9&  J135940.66+142712.0&  209.9195&  14.4533&  22.1&  21.7&  X&  X&  X&  X&  X&  X&   \\
10&  J135941.78+144035.2&  209.9241&  14.6765&  22.2&  21.9&  X&  X&  X&  X&  X&  X&   \\
11&  J135942.18+142942.2&  209.9258&  14.4951&  19.6&  19.0&  90.34&  0.38&  0.48&  5000&  -2.5&  1.5&  B\\
12&  J135943.12+144054.1&  209.9297&  14.6817&  21.2&  20.6&  -137.63&  2.38&  0.89&  6000&  -1.0&  4.5&   \\
13&  J135943.43+143438.3&  209.9310&  14.5773&  20.7&  20.3&  99.81&  1.16&  0.51&  5000&  -2.5&  1.0&  B\\
14&  J135944.57+143709.6&  209.9358&  14.6193&  20.4&  19.9&  97.93&  0.87&  0.52&  4750&  -2.5&  1.0&  B\\
15&  J135944.70+142601.8&  209.9363&  14.4338&  21.6&  20.9&  X&  X&  X&  X&  X&  X&   \\
16&  J135944.95+143230.1&  209.9373&  14.5417&  20.8&  20.3&  103.07&  1.07&  0.54&  5250&  -2.0&  2.5&  B\\
17&  J135945.06+142327.3&  209.9378&  14.3909&  21.2&  20.7&  102.84&  1.55&  0.66&  5000&  -2.0&  2.5&  B\\
18&  J135945.71+142552.5&  209.9405&  14.4313&  22.2&  21.9&  100.09&  5.87&  0.67&  5000&  -2.5&  2.0&   \\
19&  J135945.72+142230.9&  209.9405&  14.3753&  21.7&  21.2&  91.91&  2.76&  0.57&  5000&  -2.5&  4.5&   \\
20&  J135946.26+143409.2&  209.9428&  14.5692&  19.9&  19.4&  143.52&  0.55&  0.49&  5250&  -1.5&  3.0&   \\
21&  J135946.33+142511.8&  209.9431&  14.4200&  20.4&  19.9&  98.12&  0.83&  0.72&  5000&  -2.5&  2.0&  B\\
22&  J135947.06+142852.5&  209.9461&  14.4813&  21.5&  21.1&  100.99&  2.46&  0.54&  5000&  -2.0&  1.5&  B\\
23&  J135947.57+142334.5&  209.9482&  14.3929&  17.7&  16.6&  36.75&  0.15&  1.00&  4500&  -0.5&  4.0&   \\
24&  J135948.14+143646.6&  209.9506&  14.6130&  20.9&  20.3&  103.20&  1.27&  0.52&  5000&  -2.0&  1.5&  B\\
25&  J135948.33+143203.5&  209.9514&  14.5343&  19.7&  19.3&  106.17&  0.53&  0.51&  5500&  -2.0&  1.0&  B\\
26&  J135948.53+144204.4&  209.9522&  14.7012&  20.3&  19.8&  111.75&  2.08&  0.98&  5000&  -2.0&  2.0&   \\
27&  J135948.96+142428.4&  209.9540&  14.4079&  20.9&  20.4&  -34.96&  1.07&  0.52&  5250&  -1.0&  4.5&   \\
28&  J135950.13+141944.4&  209.9589&  14.3290&  20.3&  19.7&  101.74&  0.94&  0.52&  5000&  -2.0&  2.5&  B\\
29&  J135950.75+143114.2&  209.9615&  14.5206&  21.8&  21.3&  92.81&  2.58&  0.69&  5000&  -2.0&  1.5&   \\
30&  J135950.91+143002.7&  209.9621&  14.5008&  17.5&  16.5&  99.95&  0.09&  0.99&  4750&  -1.5&  1.5&   \\
31&  J135951.07+143049.8&  209.9628&  14.5138&  22.0&  21.5&  102.14&  3.18&  0.57&  4000&  -2.5&  1.0&   \\
32&  J135951.33+143905.8&  209.9639&  14.6516&  19.3&  19.3&  111.07&  1.98&  1.00&  7000&  -2.0&  2.5&   \\
33&  J135951.70+143543.3&  209.9655&  14.5954&  19.4&  18.8&  99.94&  0.30&  0.58&  4500&  -2.5&  1.0&  B\\
34&  J135951.87+143018.9&  209.9662&  14.5053&  21.3&  20.8&  35.47&  1.67&  0.53&  5000&  -1.5&  4.0&   \\
35&  J135952.11+144039.3&  209.9672&  14.6776&  21.4&  21.0&  82.79&  8.22&  1.00&  6000&  -2.5&  4.5&   \\
36&  J135952.33+143245.6&  209.9681&  14.5460&  20.7&  20.2&  99.05&  1.31&  0.52&  5000&  -2.0&  1.0&  B\\
37&  J135953.00+142232.1&  209.9709&  14.3756&  22.2&  21.8&  102.67&  5.89&  0.64&  5000&  -2.5&  3.5&   \\
38&  J135953.12+142734.3&  209.9713&  14.4595&  20.8&  20.3&  104.70&  1.28&  0.53&  5000&  -2.5&  1.5&  B\\
39&  J135953.75+143055.9&  209.9740&  14.5156&  20.7&  20.2&  122.91&  0.86&  0.50&  5000&  -1.0&  2.5&   \\
40&  J135953.93+142951.6&  209.9747&  14.4977&  22.1&  21.7&  -235.57&  4.39&  0.59&  4000&  -2.0&  1.0&   \\
41&  J135954.41+144244.2&  209.9767&  14.7123&  21.2&  20.7&  -45.51&  6.79&  1.00&  5000&  -2.5&  4.5&   \\
42&  J135954.89+143715.3&  209.9787&  14.6209&  19.5&  19.6&  109.80&  1.68&  0.55&  8000&  -2.0&  3.5&   \\
43&  J135955.33+143452.8&  209.9806&  14.5813&  19.7&  19.1&  100.03&  0.38&  0.48&  5000&  -2.0&  1.5&  B\\
44&  J135955.98+143425.6&  209.9833&  14.5738&  19.6&  19.1&  114.36&  0.38&  0.58&  5000&  -1.5&  2.5&   \\
45&  J135956.41+143556.9&  209.9851&  14.5992&  17.3&  16.3&  -37.63&  0.08&  1.00&  4750&  0.0&  4.5&   \\
46&  J135956.43+142057.4&  209.9852&  14.3493&  19.2&  18.7&  -72.15&  0.33&  0.55&  5250&  -1.0&  4.5&   \\
47&  J135956.71+142516.3&  209.9863&  14.4212&  19.8&  19.2&  13.06&  0.27&  0.47&  4750&  -1.0&  4.5&   \\
48&  J135957.84+142802.5&  209.9910&  14.4674&  20.4&  19.9&  101.64&  0.92&  0.52&  5250&  -2.0&  3.5&   \\
49&  J135958.70+144040.2&  209.9946&  14.6779&  22.3&  21.8&  107.88&  6.46&  0.59&  5000&  -2.5&  1.0&   \\
50&  J135959.70+143633.1&  209.9988&  14.6092&  20.9&  20.4&  224.82&  7.09&  0.58&  7000&  -2.5&  4.5&   \\
51&  J140000.24+143234.9&  210.0010&  14.5430&  20.6&  20.1&  108.27&  0.89&  0.56&  5000&  -2.0&  1.0&  B\\
52&  J140000.75+143529.0&  210.0031&  14.5914&  19.7&  20.0&  99.52&  2.69&  0.54&  9000&  -1.5&  3.5&   \\
53&  J140000.99+143126.7&  210.0041&  14.5241&  21.8&  21.5&  110.49&  3.12&  0.56&  6000&  -1.5&  3.0&   \\
54&  J140001.42+143424.1&  210.0059&  14.5734&  20.6&  20.1&  -28.59&  0.92&  0.50&  5500&  -2.0&  4.0&   \\
55&  J140001.53+142154.2&  210.0064&  14.3651&  19.9&  19.4&  103.50&  0.44&  0.52&  4750&  -2.5&  1.0&  B\\
56&  J140001.66+142454.8&  210.0069&  14.4152&  19.2&  18.6&  93.97&  0.26&  0.52&  5000&  -1.5&  4.5&   \\
57&  J140002.23+144114.2&  210.0093&  14.6873&  20.9&  20.3&  79.61&  1.25&  0.57&  4750&  -1.5&  2.0&   \\
58&  J140002.28+142653.4&  210.0095&  14.4482&  22.0&  21.6&  103.60&  3.60&  0.56&  5000&  -2.5&  2.5&   \\
59&  J140002.44+142249.1&  210.0102&  14.3803&  21.2&  20.6&  101.64&  1.55&  0.54&  5000&  -2.0&  3.0&  B\\
60&  J140003.07+143023.6&  210.0128&  14.5066&  21.5&  21.0&  102.06&  1.83&  0.60&  5000&  -2.5&  1.0&  B\\
61&  J140003.32+142851.4&  210.0138&  14.4810&  20.7&  20.2&  105.36&  1.67&  0.98&  4750&  -2.5&  1.5&   \\
62&  J140003.47+143952.2&  210.0145&  14.6645&  21.7&  21.3&  103.90&  4.00&  0.55&  6000&  -2.0&  3.0&   \\
63&  J140005.16+143427.8&  210.0215&  14.5744&  20.9&  20.5&  104.05&  1.39&  0.56&  5000&  -2.5&  2.0&  B\\
64&  J140005.33+143023.3&  210.0222&  14.5065&  21.0&  20.6&  100.19&  1.53&  0.52&  5000&  -2.5&  1.0&  B\\
65&  J140005.61+142618.8&  210.0234&  14.4386&  20.4&  19.9&  106.22&  0.81&  0.51&  5000&  -2.5&  1.5&  B\\
66&  J140008.67+143654.3&  210.0361&  14.6151&  20.9&  20.4&  103.07&  1.23&  0.55&  5000&  -2.5&  1.0&  B\\
67&  J140010.30+142626.5&  210.0429&  14.4407&  21.9&  21.6&  102.33&  3.48&  0.58&  5000&  -2.5&  1.0&   \\
68&  J140010.61+143823.8&  210.0442&  14.6400&  19.9&  19.4&  101.19&  0.52&  0.50&  5000&  -2.0&  2.0&  B\\
69&  J140010.69+142924.4&  210.0446&  14.4901&  19.7&  19.2&  103.55&  0.34&  0.48&  4500&  -2.5&  1.0&  B\\
\enddata

\end{deluxetable}
\begin{deluxetable}{ccccccccccccc}
\tabletypesize{\footnotesize}
\tablewidth{0pt}
\tablecaption{\label{tab:radvelctd}  }
\tablehead{ \colhead{ID}  &\colhead{Name}  &\colhead{$\alpha$}
  &\colhead{$\delta$}  &\colhead{$g$}  &\colhead{$r$}  &\colhead{RV}  
&\colhead{RV error}  &\colhead{P$_{var}$}  &\colhead{T$_{\rm eff}$}  
&\colhead{[Fe/H]}  &\colhead{$\log(g)$}  &\colhead{Bestflag}   \\ 
\\ \colhead{}  &\colhead{}  &\colhead{[deg]}  &\colhead{[deg]}
&\colhead{[mag]}  &\colhead{[mag]}  &\colhead{[$km/s$]}
&\colhead{[$km/s$]}  &\colhead{}  &\colhead{[K]}  &\colhead{}  &\colhead{}  &\colhead{}  }
\startdata
70&  J140011.53+142556.0&  210.0480&  14.4322&  21.5&  21.0&  114.41&  2.07&  0.62&  5000&  -2.5&  1.0&  B\\
71&  J140012.23+142922.0&  210.0510&  14.4894&  21.3&  20.9&  91.42&  2.61&  0.75&  6000&  -2.0&  4.5&   \\
72&  J140012.41+143327.3&  210.0517&  14.5576&  20.6&  20.3&  -90.00&  1.40&  0.52&  5000&  -2.5&  4.5&   \\
73&  J140012.92+143311.7&  210.0538&  14.5533&  19.6&  19.0&  101.39&  0.31&  0.53&  5000&  -2.0&  1.5&  B\\
74&  J140013.33+142618.1&  210.0556&  14.4384&  18.7&  18.0&  44.92&  0.16&  0.52&  4750&  -2.0&  4.5&   \\
75&  J140014.67+143930.7&  210.0611&  14.6585&  21.3&  21.0&  99.91&  2.29&  0.55&  6000&  -1.5&  4.0&   \\
76&  J140015.34+142303.0&  210.0640&  14.3842&  18.2&  17.6&  65.88&  0.12&  0.31&  5250&  -1.0&  3.5&   \\
77&  J140015.57+142348.5&  210.0649&  14.3968&  18.5&  17.7&  -53.39&  0.09&  0.37&  4750&  -0.5&  4.5&   \\
78&  J140015.81+143446.8&  210.0659&  14.5797&  21.7&  21.4&  117.33&  6.40&  0.84&  6000&  -2.5&  4.5&   \\
79&  J140016.15+143146.9&  210.0673&  14.5297&  21.2&  20.7&  X&  X&  X&  X&  X&  X&   \\
80&  J140016.59+143530.0&  210.0691&  14.5917&  19.6&  19.7&  99.39&  1.21&  0.64&  7500&  -1.0&  3.0&   \\
81&  J140016.62+142925.4&  210.0693&  14.4904&  21.3&  20.9&  97.21&  2.09&  0.55&  5000&  -2.5&  3.5&   \\
82&  J140020.58+143734.0&  210.0858&  14.6261&  22.1&  21.7&  102.01&  5.57&  0.90&  5000&  -2.5&  1.5&   \\
83&  J140021.00+143923.2&  210.0875&  14.6565&  19.5&  18.9&  -30.74&  0.32&  0.61&  4750&  -1.0&  1.5&   \\
84&  J140021.84+142553.3&  210.0910&  14.4315&  21.0&  20.5&  90.41&  1.38&  0.59&  5000&  -2.5&  1.5&  B\\
85&  J140022.10+143838.2&  210.0921&  14.6439&  19.5&  19.7&  93.84&  1.52&  0.55&  8000&  -1.5&  3.5&   \\
86&  J140022.44+143326.9&  210.0935&  14.5575&  19.6&  19.2&  98.47&  0.44&  0.49&  5250&  -2.0&  1.5&  B\\
87&  J140023.33+142607.9&  210.0972&  14.4356&  21.9&  21.3&  92.79&  2.30&  0.56&  5000&  -1.5&  1.0&   \\
88&  J140023.38+143245.2&  210.0974&  14.5459&  21.7&  21.1&  100.02&  4.23&  0.84&  5000&  -2.5&  3.0&   \\
89&  J140025.16+143346.7&  210.1048&  14.5630&  21.0&  20.6&  100.22&  1.47&  0.57&  5000&  -2.5&  2.0&  B\\
90&  J140025.49+142917.2&  210.1062&  14.4881&  21.0&  20.6&  102.16&  1.23&  0.53&  5000&  -2.5&  1.0&  B\\
91&  J140026.25+143434.4&  210.1094&  14.5762&  18.8&  18.2&  29.74&  0.17&  0.50&  5000&  -1.0&  4.5&   \\
92&  J140026.52+142919.8&  210.1105&  14.4889&  21.5&  21.2&  -117.49&  2.59&  0.60&  5000&  -2.0&  1.5&   \\
93&  J140026.57+142948.8&  210.1107&  14.4969&  21.2&  20.7&  108.10&  1.52&  0.53&  5000&  -2.5&  3.0&  B\\
94&  J140026.87+144204.2&  210.1120&  14.7012&  21.9&  21.5&  -27.44&  9.14&  1.00&  4000&  -2.5&  3.0&   \\
95&  J140027.04+143830.3&  210.1127&  14.6418&  19.6&  19.7&  105.37&  1.81&  0.55&  8000&  -2.0&  3.0&  B\\
96&  J140027.28+143219.5&  210.1137&  14.5388&  20.3&  19.8&  105.68&  0.79&  0.51&  4750&  -2.5&  1.0&  B\\
97&  J140028.13+143311.8&  210.1173&  14.5533&  22.2&  22.0&  X&  X&  X&  X&  X&  X&   \\
98&  J140028.39+142352.6&  210.1183&  14.3979&  21.4&  21.1&  98.85&  2.94&  0.65&  5000&  -2.5&  2.0&   \\
99&  J140028.73+143142.7&  210.1197&  14.5285&  21.8&  21.3&  2.25&  2.25&  0.55&  5000&  -1.5&  4.5&   \\
100&  J140028.93+142502.2&  210.1206&  14.4173&  21.8&  21.4&  103.40&  3.65&  0.58&  5000&  -2.5&  2.0&   \\
101&  J140028.95+143833.7&  210.1206&  14.6427&  22.2&  21.8&  100.54&  5.12&  0.77&  5000&  -2.5&  2.0&   \\
102&  J140031.33+143718.6&  210.1305&  14.6219&  21.9&  21.5&  -43.94&  3.64&  0.60&  5000&  -2.5&  2.0&   \\
103&  J140031.78+142015.5&  210.1324&  14.3376&  19.2&  18.7&  6.44&  0.41&  0.49&  5500&  -1.0&  4.5&   \\
104&  J140031.91+144108.3&  210.1329&  14.6856&  22.4&  22.1&  136.32&  6.38&  0.75&  4000&  -2.5&  3.5&   \\
105&  J140032.14+143627.6&  210.1339&  14.6077&  21.9&  21.4&  164.05&  3.63&  0.56&  5000&  -2.5&  3.0&   \\
106&  J140032.37+142735.8&  210.1349&  14.4600&  22.2&  21.9&  X&  X&  X&  X&  X&  X&   \\
107&  J140032.54+142400.6&  210.1356&  14.4002&  19.1&  18.5&  -60.56&  0.19&  0.76&  5250&  -0.5&  4.5&   \\
108&  J140033.07+142959.7&  210.1378&  14.4999&  19.7&  19.2&  96.86&  0.44&  0.52&  5000&  -2.5&  2.0&  B\\
109&  J140033.75+142514.3&  210.1406&  14.4206&  18.6&  17.9&  -40.08&  0.13&  0.39&  4750&  -1.0&  4.5&   \\
110&  J140035.97+142854.5&  210.1499&  14.4818&  21.6&  21.1&  -123.72&  2.82&  0.56&  5000&  -2.5&  4.5&   \\
111&  J140037.38+142858.0&  210.1558&  14.4828&  17.4&  16.4&  106.95&  0.16&  1.00&  4500&  -2.0&  1.0&   \\
112&  J140039.56+142827.8&  210.1648&  14.4744&  19.6&  19.7&  95.13&  3.71&  0.71&  8000&  -2.5&  3.5&   \\
113&  J140040.90+143208.1&  210.1704&  14.5356&  21.9&  21.7&  106.83&  4.28&  0.56&  6000&  -2.5&  4.5&   \\
114&  J140047.55+142412.0&  210.1981&  14.4033&  21.0&  20.5&  117.74&  1.62&  0.61&  5000&  -2.5&  2.5&  B\\
115&  J140047.57+142630.2&  210.1982&  14.4417&  21.3&  20.9&  99.13&  2.14&  0.57&  5000&  -2.5&  1.5&  B\\
116&  J140053.19+142705.3&  210.2217&  14.4515&  16.6&  16.1&  42.30&  0.07&  0.99&  4500&  0.0&  4.5&   \\
117&  J140057.91+142854.1&  210.2413&  14.4817&  21.7&  21.2&  103.03&  3.83&  0.57&  5000&  -2.0&  2.0&   
\enddata

\end{deluxetable}

\clearpage

\end{document}